\documentclass[aip, jcp, amsmath, amssymb, reprint, floatfix]{revtex4-1}
\usepackage[version=3]{mhchem}
\usepackage{amsmath}
\usepackage{amssymb}
\usepackage{graphicx}
\usepackage{subfigure}
\usepackage{siunitx}
\usepackage[T1]{fontenc}
\usepackage{float}
\usepackage{mathptmx}
\usepackage{xspace}
\usepackage{enumitem}
\usepackage{tabularx}
\usepackage[frozencache,cachedir=.]{minted}
\usepackage{multirow}
\usepackage{booktabs}

\usepackage{listings}
\usepackage{xcolor}
\definecolor{codegreen}{rgb}{0,0.6,0}
\definecolor{codegrey}{rgb}{0.5,0.5,0.5}
\definecolor{codepurple}{rgb}{0.58,0,0.82}
\definecolor{backcolour}{rgb}{0.95,0.95,0.92}
\definecolor{string}{RGB}{204, 112, 0}

\lstdefinelanguage{pymbe}
{
 keywords={typeof, null, catch, switch, in, int, str, float, self, import, print, def, return, from},
 keywordstyle=\color{codegreen}\bfseries,
 ndkeywords={boolean, throw, import},
 ndkeywords={class, if ,elif, endif, while, do, else, True, False , catch},
 ndkeywordstyle=\color{purple}\bfseries,
 identifierstyle=\color{black},
 sensitive=false,
 comment=[l]{\#},
 morecomment=[s]{/*}{*/},
 commentstyle=\color{codegrey}\ttfamily,
 stringstyle=\color{red}\ttfamily,
 classoffset=2,
 morekeywords= {pmb,pyMBE},
 keywordstyle=\color{blue},
 classoffset=3,
 string=[b]{"},
 stringstyle=\color{string}\ttfamily,
 morekeywords={read_cg_protein_model,
              create_protein,
              define_protein,
              center_molecule_in_simulation_box,
              define_particle,
              enable_motion_of_rigid_object,
              pymbe_library,
              units,
              set_reduced_units,
              get_reduced_units,
              setup_cpH,
              calculate_HH,
              setup_grxmc,
              setup_grxmc_unified,
              setup_gcmc,
              calculate_HH_Donnan,
              calculate_net_charge,
              load_pka_set,
              define_residue,
              define_molecule,
              filter_df,
              read_pmb_df,
              define_bond,
              add_bonds_to_espresso,
              load_interaction_parameters,
              write_pmb_df,
              to_csv,
              create_pmb_object,
              define_peptide,
              setup_lj_interactions
              },
 keywordstyle=\color{cyan},
 classoffset=4,
 morekeywords= {name, 
            filename,
            number_of_proteins, 
            topology_dict,
            residue_list,
            model,
            positions, 
            c_macro,
            molecule_name,
            particle_pairs,
            bond_parameters,
            salt_cation_name,
            salt_anion_name,
            activity_coefficient,
            z,
            diameter,
            acidity,
            pka,
            pka_set,
            epsilon,
            espresso_system,
            unit_length,
            unit_charge,
            temperature,
            counter_ion,
            constant_pH,
            SEED,
            object_name,
            pH_list,
            pH_res,
            c_salt_res,
            proton_name,
            hydroxide_name,
            sodium_name,
            chloride_name,
            object_names,
            c_salt,
            central_bead,
            side_chains,
            pmb_type,
            sequence,
            excess_chemical_potential,
            molecule_id,
            residue_id,
            particle_id,
            state_one,
            state_two,
            es_type,
            label, 
            bond_object,
            bond_type,
            particle_name1,
            particle_name2,
            charge,
            position,
            cutoff,
            number_of_objects,
            particle_id1,
            particle_id2,
            bond_object,
            sigma,
            offset,
            },
 keywordstyle=\color{purple},
 classoffset=5,
 morekeywords= {particle_id, 
            },
 keywordstyle=\color{violet},
}

\lstdefinelanguage{python3}
{
 keywords={typeof, null, catch, switch, in, int, str, float, self, import, print},
 keywordstyle=\color{codegreen}\bfseries,
 ndkeywords={boolean, throw, import},
 ndkeywords={return, class, if ,elif, endif, while, do, else, True, False , catch, def, from},
 ndkeywordstyle=\color{purple}\bfseries,
 identifierstyle=\color{black},
 sensitive=false,
 comment=[l]{\#},
 morecomment=[s]{/*}{*/},
 commentstyle=\color{codegrey}\ttfamily,
 stringstyle=\color{red}\ttfamily,
 classoffset=2,
 morekeywords= {--filename,filename }
 keywordstyle=\color{cyan},
}

\lstdefinestyle{mystyle}{
    language=python,
    commentstyle=\color{codegreen},
    keywordstyle=\color{magenta},
    numberstyle=\tiny\color{codegray},
    stringstyle=\color{codepurple},
    basicstyle=\ttfamily,
    breakatwhitespace=false,         
    breaklines=true,                 
    captionpos=b,                    
    keepspaces=true,                 
    numbersep=5pt,                  
    showspaces=false,                
    showstringspaces=false,
    showtabs=false,                  
    tabsize=2,
}

\newcommand{\code}[1]{{\lstinline[language=pymbe] !#1!}}

\usepackage[textsize = small]{todonotes}

\usepackage[english]{babel}
\usepackage[colorlinks=true,
            linkcolor=blue,
            urlcolor=blue,
            citecolor=blue]{hyperref}
\usepackage{multirow, makecell}
\usepackage{xcolor,colortbl}
\graphicspath{{./figures/}}

\newcommand*{\etal}{\emph{et al. \xspace}}
\newcommand*{\ie}{\emph{i.e. \xspace}}

\newcommand*{\refeq}[1]{Eq.~\ref{#1}}
\newcommand*{\kT}{k_\mathrm{B}T}

\newcommand*{\muref}{\mu^{\ominus}}

\newcommand*{\muex}{\mu^\mathrm{ex}}
\newcommand*{\mures}{\mu^\mathrm{res}}
\newcommand*{\Ires}{I^\mathrm{res}}

\newcommand*{\pK}{\mathrm{p}K}
\newcommand*{\pKa}{\pK_{\mathrm{A}}}

\newcommand*{\pKai}{\pK_{\mathrm{A}}^{\mathrm{i}}}
\newcommand*{\Kai}{K_{\mathrm{A}}^{\mathrm{i}}}

\newcommand*{\Kw}{K_{\mathrm{w}}}

\newcommand*{\pH}{\mathrm{pH}}
\newcommand*{\espresso}{ESPResSo\xspace}
\newcommand*{\pymbe}{pyMBE\xspace}
\newcommand*{\ideq}{\overset{\mathrm{ideal}}{=}}
\newcommand*{\grxmc}{G-RxMC\xspace}
\newcommand*{\cref}{c^{\ominus}}
\newcommand*{\cres}{c^{\mathrm{res}}}

\newcommand{\mycomment}[1]{}

\usepackage{hyperref}

\makeatletter
\def\@email#1#2{%
 \endgroup
 \patchcmd{\titleblock@produce}
  {\frontmatter@RRAPformat}
  {\frontmatter@RRAPformat{\produce@RRAP{*#1\href{mailto:#2}{#2}}}\frontmatter@RRAPformat}
  {}{}
}%
\makeatother

\begin{document}

\preprint{APS/123-QED}

% Title
\title{pyMBE: the Python-based Molecule Builder for ESPResSo}
% Affiliations
\newcommand*{\affCUNI}{Department of Physical and Macromolecular Chemistry, Faculty of Science, Charles University, Hlavova 8, 128 40 Prague 2, Czech Republic}
\newcommand*{\affUTN}{Grupo de Bionanotecnologia y Sistemas Complejos. Infap-CONICET \& Facultad Regional San Rafael, Universidad Tecnológica Nacional, 5600 San Rafael, Argentina}
\newcommand*{\affIQTCUB}{Department of Material Science and Physical Chemistry, Research Institute of Theoretical and Computational Chemistry (IQTCUB), University of Barcelona, Martí i Franquès 1, 08028 Barcelona, Spain}
\newcommand*{\affSTU}{Institute for Computational Physics, University of Stuttgart, Allmandring 3, 70569 Stuttgart, Germany}
\newcommand*{\affNTNU}{Department of Physics, NTNU - Norwegian University of Science and Technology, NO-7491 Trondheim, Norway}

% Authors

\author{David Beyer}
\thanks{These two authors contributed equally.}
\affiliation{\affSTU}

\author{Paola B. Torres}
\thanks{These two authors contributed equally.}
\affiliation{\affUTN}

\author{Sebastian P. Pineda}
\affiliation{\affCUNI}

\author{Claudio F. Narambuena}
\affiliation{\affUTN}

\author{Jean-Noël Grad}
\affiliation{\affSTU}

\author{Peter Košovan}
\affiliation{\affCUNI}

\author{Pablo M. Blanco}
\email{pablb@ntnu.no}
\affiliation{\affCUNI}
\affiliation{\affIQTCUB}
\affiliation{\affNTNU}

% Keywords

% Synopsis (max 200 words)

\date{\today}

\begin{abstract}
We present the Python-based Molecule Builder for ESPResSo (\pymbe), an open source software to design custom Coarse-Grained (CG) models, as well as pre-defined models of polyelectrolytes, peptides and globular proteins in the Extensible Simulation Package for Research on Soft Matter (\espresso).
The Python interface of \espresso  offers a flexible framework, capable of building custom CG models from scratch.
As a downside, building CG models from scratch is prone to mistakes, especially for newcomers in the field of CG modeling, or for molecules with complex architectures.
The \pymbe module builds CG models in \espresso using a hierarchical bottom-up approach, providing a robust tool to automate the setup of CG models and helping new users prevent common mistakes.
\espresso features  the constant pH (cpH) and grand-reaction (\grxmc) methods, which have been designed to study chemical reaction equilibria in macromolecular systems with many reactive species.
However, setting up these methods for systems which contain several types of reactive groups is an error-prone task, especially for beginners. 
The \pymbe module enables the automatic setup of cpH and \grxmc simulations in \espresso, lowering the barrier for newcomers and opening the door to investigate complex systems not studied with these methods yet.
To demonstrate some of the applications of \pymbe, we showcase several case studies where we successfully reproduce previously published simulations of charge-regulating peptides and globular proteins in bulk solution and weak polyelectrolytes in dialysis. 
The \pymbe module is publicly available as a GitHub repository (\href{https://github.com/pyMBE-dev/pyMBE}{https://github.com/pyMBE-dev/pyMBE}) which includes its source code and various sample and test scripts, including the ones that we used to generate the data presented in this article.
\end{abstract}

\maketitle

\section{Introduction}
Computer simulations using molecular models are a valuable tool to rationalize experimental observations, validate the applicability of simplified theories and provide numerical results beyond the capabilities of analytical calculations.
The most popular molecular models are atomistic models, in which each atom is represented by an explicit particle. 
While atomistic models provide a detailed description of a system, they are often limited to nanoscopic scales due to the current computing capabilities. 
Coarse-graining techniques circumvent this problem by reducing the number of degrees of freedom of a system, at the cost of reducing the model resolution. 
Coarse-grained (CG) models are often preferred to investigate macromolecular systems, because they allow to reach longer time and length scales than atomistic models. 
Despite the popularity of CG models to simulate macromolecular systems, most of the current software packages for molecular modelling are designed for atomistic models. 
Among software packages that natively support CG models, remarkable examples are the Extensible Simulation Package for Research on Soft Matter (\espresso),\cite{weik19a,weeber24a} ESPResSo++,\cite{guzman19a} Faunus,\cite{lund2008a,stenqvist2013a} FEASST,\cite{hatch2018a} GROMACS,\cite{abraham2015a} HOOMD-blue,\cite{anderson20a} the modular molecular simulation (MOLSIM)\cite{rescic15a} and the Large-scale Atomic/Molecular Massively Parallel Simulator (LAMMPS).\cite{thompson22a}

Only a few software packages natively support CG models, such as the Extensible Simulation Package for Research on Soft Matter (\espresso),\cite{weik19a,weeber24a} ESPResSo++,\cite{guzman19a} HOOMD-blue,\cite{anderson20a} the modular molecular simulation (MOLSIM)\cite{rescic15a} and the Large-scale Atomic/Molecular Massively Parallel Simulator (LAMMPS).\cite{thompson22a}

\begin{figure}
\centering
\includegraphics[width=0.5\columnwidth]{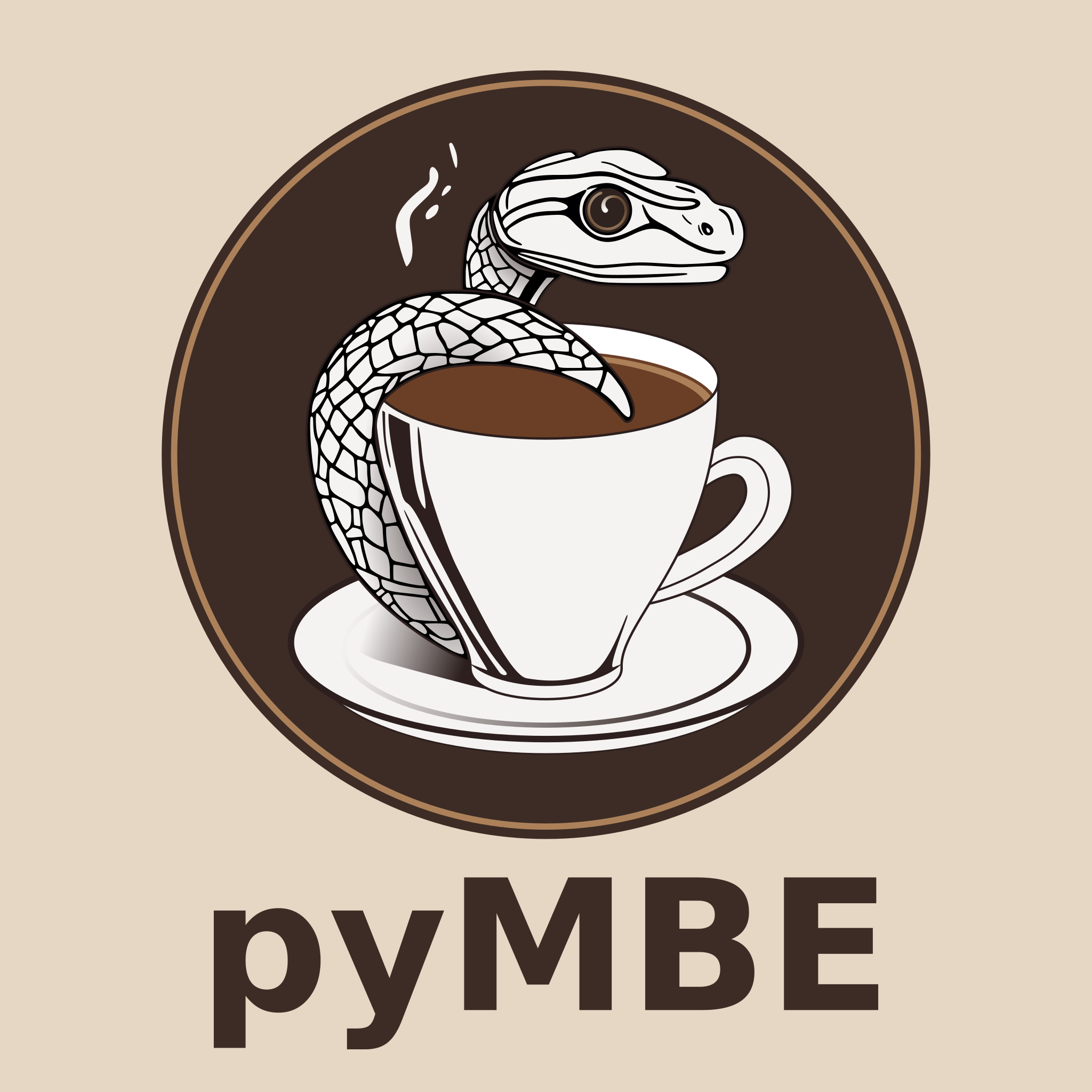}
\caption{\label{fig:logo} Logo of the Python-based Molecule Builder for \espresso (\href{https://github.com/pyMBE-dev/pyMBE}{pyMBE}). The logo has been edited from an image originally generated using Artificial Intelligence powered tools from Canva.\cite{canva2023}}
\end{figure}

CG models are also well-suited for Monte Carlo methods, which can be used to sample reversible chemical reactions in reactive systems. 
These reactions are usually not considered in molecular simulations.
Nevertheless, they can significantly affect the properties of various macromolecular systems by affecting their ionization states.\cite{lund13a, landsgesell19a, blanco23a}
Constant pH Monte Carlo\cite{reed92a} (cpH) simulations are a popular tool to study pH-sensitive molecules in bulk solution, such as solutions of weak polyelectrolytes,\cite{laguecir06a, ullner96b, ullner96a} peptides,\cite{blanco21a, lunkad21b, lunkad22b, pineda2024} proteins,\cite{lund05a, dasilva09a,lund13a, torres19a,torres22a} polyelectrolyte microgels\cite{hoare04a, hofzumahaus18a, hofzumahaus21a, strauch23a} and charged nanoparticles.\cite{clavier15a, stornes21a} 
In contrast, the grand-reaction (\grxmc) method has been designed to study two-phase systems,\cite{landsgesell20b,landsgesell20b-err}  termed "system" and "reservoir".
Macromolecules or nanoparticles are present only in the system and cannot be exchanged with the reservoir.
However, small molecules and ions can be exchanged between the system and reservoir.
Examples of such two-phase systems include weak polyelectrolyte hydrogels,\cite{ricka84a, tang20a, philippova97a, landsgesell22a, beyer22a} coacervates,\cite{love20a, stano23a} brushes,\cite{delcastillo20a, delcastillo20b, yuan21a, beyer23a, yuan23a, balzer23a} and solutions of polyelectrolyte chains\cite{landsgesell20b} or proteins.\cite{briskot22a}
The cpH and the \grxmc methods are available in \espresso, however, setting them up for systems with many different reaction types often leads to repetitive and lengthy scripts, prone to errors which may be difficult to identify.
Therefore, a tool to automate the setup of these methods is highly desirable.

The first step before performing a computer simulation is building the molecular model in the software. 
Many molecule builders such as Avogadro,\cite{hanwell12a}
the TopoTools plugin for Visual Molecular Dynamics\cite{humphrey96a} (VMD),
the Builder tool in PyMOL,\cite{pymol23a}
the LEaP molecular builder\cite{salomonferrer13a} 
from the AmberTools suite,\cite{case23a}
the mBuild\cite{sallai14a,klein16a} 
tool in MoSDeF,\cite{cummings21a}
PySIMM\cite{demidov2022},
pyPolyBuilder\cite{ramos21a}
or Atlas\cite{atlas23a} 
are available to facilitate the setup of atomistic models in software for molecular simulation.
While some of these all-atom molecule builders can be repurposed for CG modeling by defining custom molecular residue types, this procedure is time-consuming and requires detailed knowledge of the internal representation of the residues in the software.

Unlike all-atom models, CG models are not uniquely defined by the molecular structure.
Therefore, they are often problem-specific, and providing a universal builder for CG models is not as straightforward as for all-atom models.
Consequently, many users design their own scripts to build CG models for a particular simulation software, which can consume a significant amount of their working time.
Building CG models from scratch is susceptible to errors and often leads to repeating common mistakes in the simulation setup, especially by new users of a software.

To bypass these shortcomings of setting up CG models, there is a growing number of tools to automate the coarse-graining of existing all-atom models: martinize.py,\cite{dejong12a} Martinize2,\cite{kroon22a} Auto\_Martini,\cite{bereau15b} MAD,\cite{hilpert23a} \mbox{PDB$\to$UNF} Converter,\cite{kutak21a} GENESIS-CG-tool.\cite{tan22a}
There are also some special-purpose CG molecule builders for example for DNA origami\cite{doye13a,majikes21a} and lipid membranes.\cite{wassenaar15a}
However, general-purpose CG molecule builders capable of generating arbitrary polymer topologies are still very scarce.
Currently, the only molecule builders with such capabilities are Moltemplate\cite{jewett21a} and MoSDeF-GOMC.\cite{crawford23a}
Moltemplate is designed to facilitate the setup of input files for LAMMPS whereas MoSDeF-GOMC is a Python interface for GOMC,\cite{nejahi2021a} designed to set up Monte Carlo simulations of CG models of small molecules and crystals in various statistical ensembles.

\espresso offers a flexible Python scripting interface capable of building custom CG models from scratch, resulting in a convenient software to perform computer simulations of CG models.
However, building CG models from scratch is an error-prone and tedious task for beginners, especially for molecules with complex structures.
In practice, this has prevented some users of \espresso from using the software to simulate models with complex architectures or force-fields.
Although Moltemplate can be used with a deprecated \espresso v3.3.1, none of the above-mentioned CG molecule builders is compatible with the current version of \espresso.

In this article, we present the Python-based Molecule Builder for \espresso (\pymbe),  an open-source software developed as a collaborative project between various research groups modeling weak polyelectrolytes and biomacromolecules which require rather complex model setups, including acid-base reactions and chemical equilibria in two-phase systems.
The \pymbe module is a molecule builder that has been designed to build CG models of macromolecules with different complex architectures, ranging from flexible polyelectrolytes and peptides to globular proteins with a rigid structure.
In addition, it facilitates the setup of the cpH and \grxmc methods in \espresso, lowering the barrier to entry for new users of these methods.
In Fig. \ref{fig:logo}, we present the logo of \pymbe with its mascot: a python snake that loves drinking espresso while coding.
To demonstrate the applications of the \pymbe module, we showcase examples where it successfully reproduces reference data from previous works done with \espresso and other simulation software. 
Altogether, \pymbe aims to aid researchers in the field to ensure both reproducibility and reusability of their simulations. 

\begin{table}[H]
\renewcommand{\arraystretch}{1.5}
\begin{tabular}{|p{.98\columnwidth}|}
 \hline
  \rowcolor{cyan!50!white}[\tabcolsep] \multicolumn{1}{|c|}{Table I: Features of pyMBE}  \\ 
 \hline
  $\bullet$ Molecule builder designed for \espresso. \\
  $\bullet$ Build custom polymer molecules using simple blocks.   \\
  $\bullet$ Support for coarse-grained models of peptides and proteins. \\
  $\bullet$ Automated setup of cpH and \grxmc simulations. \\
  $\bullet$ Bookkeeping of the topology of the molecules. \\
 \hline
\end{tabular}
\refstepcounter{table}\label{tab:features}
\end{table}

\section{Features of pyMBE}
\pymbe is distributed under the GPL-3.0 license, that guarantees its users to freely run, study, share and modify the software.
The \pymbe module has been designed as a toolbox to aid the setup of simulations in \espresso.
Therefore, the standard use of \pymbe is to be imported as a library together with \espresso in Python scripts for molecular simulation.
However, the functionalities of \pymbe as a molecule builder for coarse-grained models are not restricted to \espresso and they could be extended to other software for molecular simulation.
The \pymbe module is under active development and new functionalities are developed depending on the needs of the community.
Following the "Four-eyes" principle, every new functionality is reviewed by one or more members of the \pymbe development team before being merged into the stable version of \pymbe.
The stable version of \pymbe is automatically periodically tested to reproduce previous reference data from our groups.
\pymbe is an open community and we welcome new users and developers to join the project and contribute to its development.
The \pymbe module is currently under active development and new functionalities are still being implemented by the development team.
Before being merged into the stable version of \pymbe, every new functionality is reviewed by one or more members of the development team and tested to reproduce previous reference data from our groups. 
A stable version of \pymbe is publicly available as a GitHub repository (\href{https://github.com/pyMBE-dev/pyMBE}{https://github.com/pyMBE-dev/pyMBE}), including the source code, a documentation of the Application Programming Interface (API), a Jupyter notebook tutorial for beginners and several sample scripts and test scripts. 
The API documentation is automatically generated from the source code of the library using the Python library pdoc\cite{pdoc} and it can be consulted in the GitHub website of \pymbe (\href{https://pymbe-dev.github.io/pyMBE}{https://pymbe-dev.github.io/pyMBE}).
Currently, \pymbe uses various Python libraries, listed in Appendix \ref{sec:dependencies}.
However, we note that \pymbe is a live project, so the readers should consult the project repository for an up-to-date list of dependencies and documentation.
We outline in Table \ref{tab:features} the main features of \pymbe, which we will explain throughout this section.
\subsection{\label{sec:basics} Basic setup and units} 
The first step when using \pymbe is to import it as a Python library,
\begin{lstlisting}[language=pymbe,linewidth=\columnwidth,breaklines=true,caption={Import pyMBE and create an instance of the library.},label={lst:init}] 
import pyMBE
pmb = pyMBE.pymbe_library(seed = 42)
\end{lstlisting}
which enables its use with standard \espresso scripts. 
When creating an instance of the library, it is mandatory to provide a \code{seed} for the various random number generators.
\pymbe employs the Pint\cite{pint} library to facilitate the work with physical quantities and units.
When the user creates an instance of \pymbe, it creates an instance of a UnitRegistry object from Pint, which is stored as an attribute of \pymbe. The user can access this attribute to perform operations with units, as conventionally done when using Pint: 
\begin{lstlisting}[language=pymbe,linewidth=\columnwidth,breaklines=true, caption={Defining physical quantities with units.},label={lst:units}]
k_B = 1.38e-23*pmb.units.J/pmb.units.K
T   = pmb.units.Quantity(298.15, "K")
kT  = k_B*T
\end{lstlisting}
Here, \code{k_B}, \code{T} and \code{kT} are \code{pint.Quantity objects}, storing information about the magnitude and the units of each variable.
When operating with \code{pint.Quantity objects}, Pint internally handles  unit conversion and it resolves the resulting dimensionality. 
For example, in Code snippet \ref{lst:units}, Pint assigns the unit of Joule to \code{kT} when it executes the operation \code{k_B*T}:
\begin{lstlisting}[language=pymbe,linewidth=\columnwidth,breaklines=true, caption={Printing a pint.Quantity object.},label={lst:print_units}]
>>> print(kT)
4.11447e-21 joule
\end{lstlisting}

The use of Pint permits an easy and reliable transformation from the international system of units to the reduced system of units used in \espresso.
By default, \pymbe uses a set of internal units, based on the commonly used set of reduced units in simulations of Lennard-Jones systems.\cite{frenkel-smit}
In this system of units, energies are divided by the thermal energy $\kT$, distances are divided by the particle diameter, $\sigma$, and charges are divided by the elementary charge $e$.

In \pymbe,  the internal unit of energy is the thermal energy $\kT$, and the default value of temperature is $T = \qty{298.15}{K}$.
The internal unit of charge is the elementary charge $e = 1.602\cdot10^{-19}\,\qty{}{C}$ and the internal unit of length is $\sigma = \qty{0.355}{nm}$.
Additionally, the strength of electrostatic interactions is set by the Bjerrum length, $l_{\mathrm{B}} \approx \qty{0.71}{nm} = 2\sigma$, which corresponds to aqueous solutions at $T = \qty{298.15}{K}$.
This ratio between particle size and Bjerrum length ensures that activity coefficients (excess chemical potentials) of simple electrolyte solutions, modeled as Lennard-Jones  spheres in implicit solvent, approximately match the experimental activity coefficients of aqueous solutions of NaCl.\cite{landsgesell20b}
The same value has been commonly used in other simulations by some of the authors of the current article.
The values of various quantities in the system of reduced units used by \espresso are then obtained by dividing their values in SI units by the internal units.
For example, if the desired box length is $L=\qty{3.55}{nm}$, then the value of the reduced box length will be $l = \qty{3.55}{nm} / \qty{0.355}{nm} = 10$.
The above choice of internal units is by no means universal and a different choice may be preferred for some purposes.
Therefore, user can choose a different set of internal units in \pymbe and print the active set of reduced units in the library:
\begin{lstlisting}[language=pymbe,linewidth=\columnwidth,breaklines=true,caption={Setting up a custom set of internal units and printing it to screen.},label={lst:custom-units}]
pmb.set_reduced_units(
    unit_length = 0.5*pmb.units.nm,  
    unit_charge = 2*pmb.units.e, 
    temperature = 300*pmb.units.K)
print(pmb.get_reduced_units())
\end{lstlisting}
By default, there is no pre-defined internal unit of mass in \pymbe, because this choice does not influence the equilibrium properties of the system in the absence of an external gravitational field, which is a common case in molecular modeling.
The conversion between the system of reduced units used by \espresso and the SI units in which the users usually obtain their input parameters is a common problem which leads to mistakes in the simulation setup.
The \pymbe module alleviates this issue by allowing the user to provide the inputs in SI units, and internally ensures that the conversion to reduced units is done correctly.

\subsection{\label{sec:def_part}Defining particles, residues and molecules}

\subsubsection{Particles}
\begin{figure}
\centering
\includegraphics[width=\columnwidth]{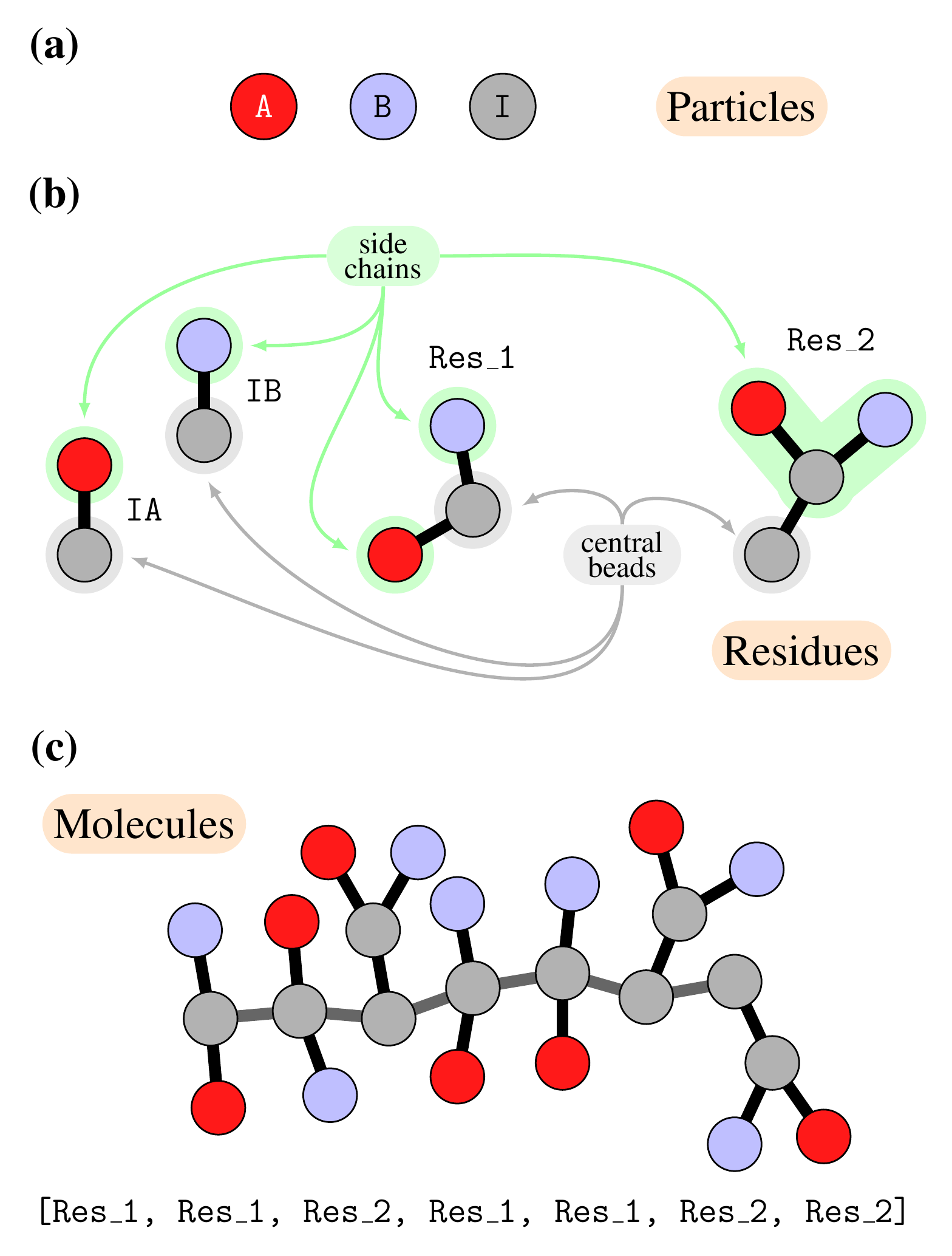}
\caption{\label{fig:scheme_building_blocks} Schematic representation of the essential building blocks used in \pymbe: particles, residues and molecules. (a): A particle is the smallest building block in the \pymbe module, representing an atom or a coarse-grained group of atoms. (b): Residues consist of a central particle, referred as central bead, with one or multiple side chains bonded to it. (c): Molecules are built as a linear sequence of residues with an arbitrary user-defined composition.
}
\end{figure}

The \pymbe module uses a hierarchical object-oriented set of fundamental building blocks: particles, residues and molecules, as schematically shown in Fig. \ref{fig:scheme_building_blocks}. 
This hierarchical approach is partly inspired by the architecture used in the MOLSIM software.\cite{rescic15a}  
The smallest building block in \pymbe is called a particle. 
It can represent an atom or a coarse-grained group of atoms. 
Users can use \pymbe to define the properties of various particle types:   
\begin{lstlisting}[language=pymbe,linewidth=\columnwidth,breaklines=False, caption={Defining particle properties.}, label={lst:define-particle}]
# Inert particle 
pmb.define_particle(
    name = "I",
    sigma = 0.3*pmb.units.nm,
    epsilon = kT,
    z = 0)
    
# Acidic particle
pmb.define_particle(
    name = "A",
    acidity = "acidic",
    pka = 4)
    
# Basic particle
pmb.define_particle(
    name = "B",
    acidity = "basic",
    pka = 9)
\end{lstlisting}
The only required argument \code{name} is used as an identifier by which the user later refers to this particle type when building residues and molecules, setting up their interactions and chemical reactions. 
All other arguments are optional.
The arguments \code{sigma} and \code{epsilon} are used in the setup of pairwise Lennard-Jones (LJ) interactions.
Additionally, the user can provide the arguments \code{cutoff} and \code{offset} that allow further tuning of the LJ interactions. 
We provide a detailed explanation of these LJ-related parameters in Section \ref{sec:setup}.
The optional argument \code{z} corresponds to the charge number of the particle and thus sets a permanent charge $q=ze$ of a particle.
Alternatively, the user can specify \code{acidity} of the particle, which can be \code{"acidic"} or \code{"basic"}.
In such a case, the particle may have two states: one neutral and one with $z=-1$ if it is acidic or $z=+1$ if it is basic.
When setting the \code{acidity}, the user must also specify the value of the acidity constant, \code{pka}.
The default value of the \code{acidity} is \code{None}, which means that the particle has only one state with a permanent charge \code{q}.
The \code{acidity} and \code{pKa} are used when setting up the constant-pH or the grand-reaction method, as explained in Section \ref{sec:methods}.

\subsubsection{Residues}
Residues in \pymbe are composed of one or multiple particles and they are used as monomeric units to build molecules. 
A residue is defined as a central particle to which other particles are connected, as shown in Fig. \ref{fig:scheme_building_blocks} (b):
\begin{lstlisting}[language=pymbe,linewidth=\columnwidth,caption={Defining simple residues.}, label={simple-residues}]
# Acidic residue
pmb.define_residue(
    name = "IA",
    central_bead = "I",
    side_chains = ["A"])

# Basic residue
pmb.define_residue(
    name = "IB",
    central_bead = "I",
    side_chains = ["B"])
\end{lstlisting}
More complex residues can be created by using multiple particles or residues as side chains, also shown in Fig. \ref{fig:scheme_building_blocks} (b):
\begin{lstlisting}[language=pymbe,linewidth=\columnwidth,caption={Defining complex residues.}, label={complex-residues}]
pmb.define_residue(
    name = "Res_1",
    central_bead = "I",
    side_chains = ["A","B"])

pmb.define_residue(
    name = "Res_2",
    central_bead = "I",
    side_chains = ["Res_1"])    
\end{lstlisting}
Here, \code{name} is the identifier of the residue. 
The \code{central_bead} should be the name of a particle and \code{side_chains} should be a list of particles or residues defined in \pymbe.
The names listed in \code{side_chains} determine which objects will be bonded to \code{central_bead}.
These names must correspond to particle or residue objects which have been previously defined.
If the user provides the name of a residue object, \pymbe defines that the \code{central_bead} of that residue will be bonded to the \code{central_bead} of the defined residue.

\subsubsection{\label{sec:custom} Custom molecules}
\begin{figure}
\centering
\includegraphics[width=\columnwidth]{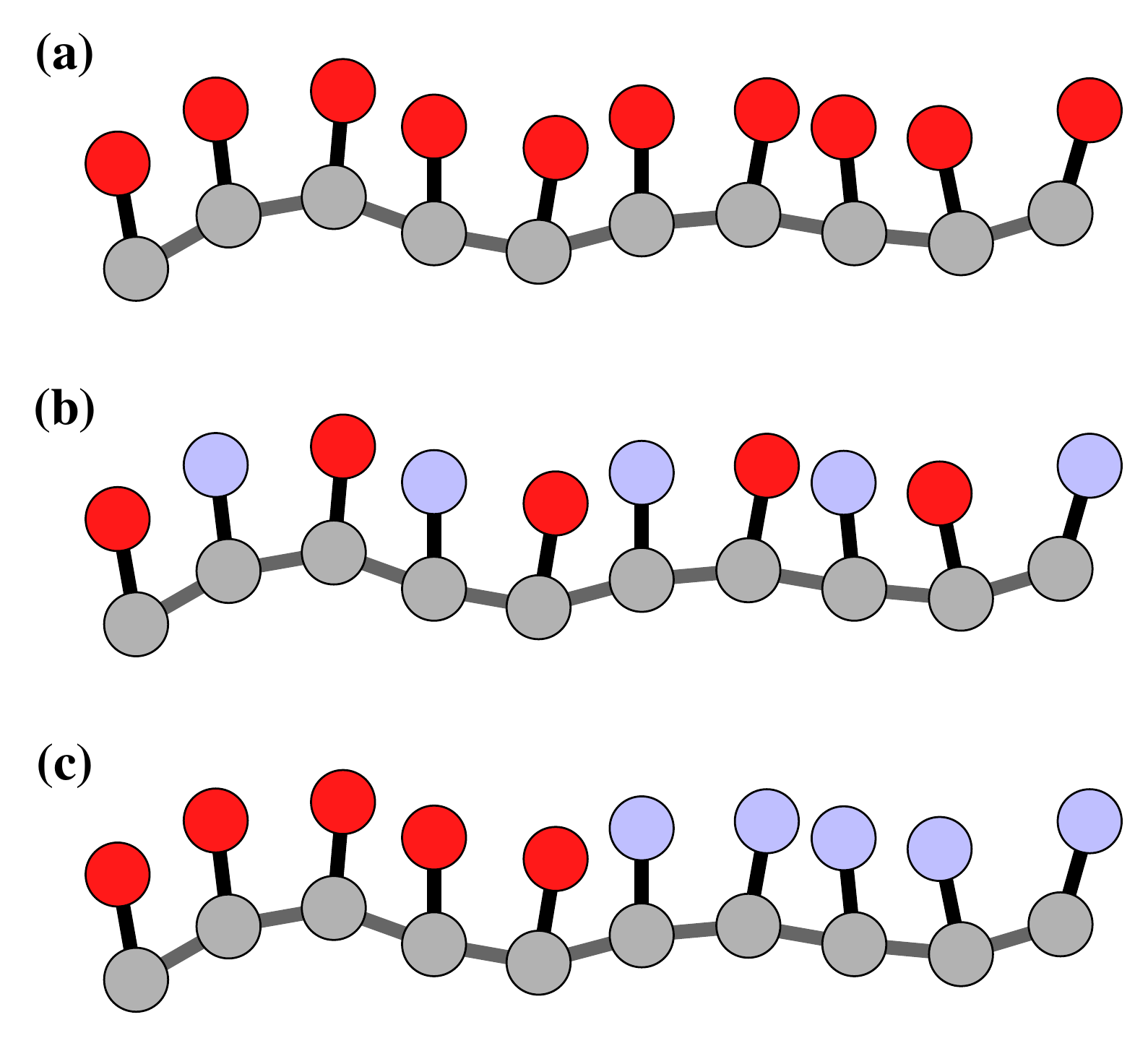}
\caption{\label{fig:scheme_list_chains} Schematic representation of various chain molecules which can be created using \pymbe. (a): A polyacid. (b): An alternating polyampholyte copolymer. (c): A diblock polyampholyte copolymer.}
\end{figure}

Custom chain molecules are built as sequences of residues, as shown in Fig. \ref{fig:scheme_building_blocks} (c):
\begin{lstlisting}[language=pymbe,linewidth=\columnwidth,caption={Defining a molecule.}, label={chain-molecule}]
pmb.define_molecule(
    name = "A_molecule",
    residue_list = ["Res_1", "Res_1",
                    "Res_2", "Res_1",
                    "Res_1", "Res_2",
                    "Res_2"])
\end{lstlisting}
Here, \code{name} is the identifier of the molecule, composed of residues specified in \code{residue_list}.
The sequence of residues in the list defines the order in which they are connected to form the molecule.
Chain molecules with multiple repeating units of the same type can be conveniently created with \pymbe using the built-in Python functionality for creating lists with repeating elements, for example (cf. Fig. \ref{fig:scheme_list_chains}):
\begin{lstlisting}[language=pymbe,linewidth=\columnwidth,caption={Defining various chain arquitectures.}, label={lst:list-molecules}]
# Polyacid
pmb.define_molecule(
    name = "polyacid",
    residue_list = ["IA"]*10)
    
# Alternating polyampholyte
pmb.define_molecule(
    name = "alternating_polymer",
    residue_list = ["IA","IB"]*5)
    
# Diblock polyampholyte
pmb.define_molecule(
    name = "diblock_polymer",
    residue_list = ["IA"]*5+["IB"]*5)
\end{lstlisting}
In principle, \pymbe could be used also to create more complex branched molecules, however, it has been primarily designed to create linear chain molecules with short side chains.
A generalization to conveniently create molecules with arbitrary architectures is planned for the future.

\subsubsection{\label{sec:file-load}Loading parameters from a file}
As an alternative to defining each particle and residue in the Python script, \pymbe also features a helper function to load these parameters from a file:
\begin{lstlisting}[language=pymbe,linewidth=\columnwidth,caption={Loading interaction parameters.}]
pmb.load_interaction_parameters(
    filename = "path_to_file") 
\end{lstlisting}
where \code{filename} contains a path to the local file, which should be structured as a JSON dictionary:
\begin{lstlisting}[language=pymbe,linewidth=\columnwidth,caption={General parametrization in JSON format.}]
{"object_type":"particle", 
 "name": "CA", 
 "z":0, 
 "sigma": {"value":0.35, "units":"nm"}}
\end{lstlisting}
More complete examples of such files can be found in the \pymbe repository, folder `\texttt{parameters/peptides}'. 
A similar helper function is available to load the \code{acidity} and \code{pka} values for a set of particles 
\begin{lstlisting}[language=pymbe,linewidth=\columnwidth, caption={Loading acid-base properties.}]
pmb.load_pka_set(
    filename = "path_to_pka_set")
\end{lstlisting}
where the input file has the following structure:
\begin{lstlisting}[language=pymbe,linewidth=\columnwidth,caption={Acid-base properties in JSON format.}]
{"D" : {"pka_value": 4.0, 
        "acidity": "acidic"}}
\end{lstlisting}
In the \pymbe repository, we provide some specific sets of $\pKa$ values for amino acid residues, based on well established  compilations in the literature: 
Bienkiewicz and Lumb,\cite{bienkiewicz1999} 
CRC Handbook of Chemistry and Physics,\cite{lide1991a} 
Hass and Mulder,\cite{hass2015} 
Nozaki and Tanford,\cite{nozaki1967a} 
Platzer \etal\cite{platzer2014} and 
Thurlkill \etal\cite{thurlkill2006}
These pre-defined sets can be found in the \pymbe repository in the folder `\texttt{parameters/pka\_sets}'.
In addition to the above, some less common sets $\pKa$ values are included in the repository.
These sets refer to studies which we used for benchmarking the results produced by \pymbe.
They are required for internal purposes, namely for integration and system testing of \pymbe within the continuous integration framework.

\subsubsection{\label{sec:pre-defined}Pre-defined molecules}
\begin{figure}
\centering
\includegraphics[width=\columnwidth]{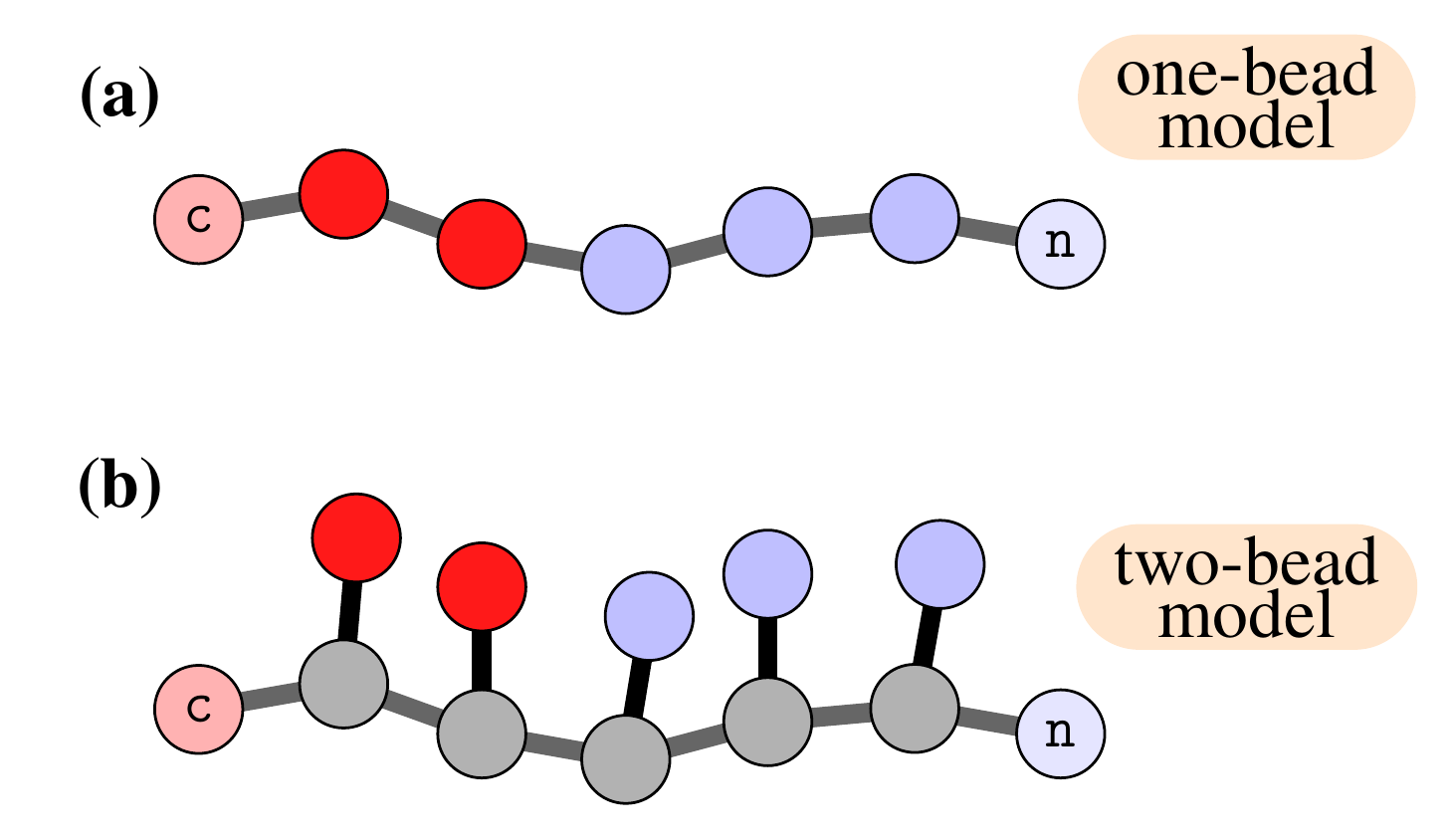}
\caption{\label{fig:scheme_peptides} Schematic representation of the peptide $\mathrm{Cys}_2\mathrm{His}_3$ using two pre-defined coarse-grained models: a one-bead representation (a) and a two-bead representation (b).}
\end{figure}

In addition to building custom molecules defined by the user, the \pymbe module also provides specific functions to build pre-defined coarse-grained models of flexible peptides and rigid proteins. 
To begin with, it is necessary to load a set of particle parameters and $\pKa$ values for amino acids from the \pymbe repository, as described in the previous section.
Then, peptides can be defined in a way similar to custom molecules:
\begin{lstlisting}[language=pymbe,linewidth=\columnwidth,caption={Defining a peptide.}]
pmb.define_peptide(
    name = "Cys2His3", 
    sequence = "cCCHHHn", 
    model = "2beadAA")
\end{lstlisting}
where \code{name} is again the identifier of the peptide molecule. 
The argument \code{sequence} uses the standard one-letter or three-letter codes for amino acids.\cite{iupaciub84a}
For example, valid \code{sequence} inputs for the peptide $\mathrm{Cys}_2\mathrm{His}_3$ are \code{"CCHHH"}, \code{"Cys-Cys-His-His-His"} or \code{"CYS-CYS-HIS-HIS-HIS"}.
The small letters \code{"c"} and \code{"n"} in the peptide sequence specify that an additional residue should be added at the end of the molecule, corresponding to the amine group at the N-terminus and carboxylate group at the C-terminus of the peptide.
If these are not explicitly specified, then the C-terminal and N-terminal groups are not added.
The function \code{define_peptide} internally defines a residue object in \pymbe for each type of amino acid in \code{sequence}.
The argument \code{model} defines how individual amino acids are represented as residues.
In a one-bead CG model (cf. Fig. \ref{fig:scheme_peptides} (a)), \code{model = "1beadAA"}, each amino acid residue is represented by a single particle.
In a two-bead CG model  (cf. Fig. \ref{fig:scheme_peptides} (b)), \code{model = "2beadAA"}, each amino acid residue is represented by two particles.
These particles are internally identified in \pymbe as \code{name = "CA"} for the central bead (same for all residue types) and a side-chain particle with the same \code{name} given by the corresponding one-letter code of the amino acid.
Exceptions to this rule are Glycine and the \code{"c"} and \code{"n"} end group residues, which are always represented by a single particle, named by the respective one-letter code. 
Although \code{sequence} uses a nomenclature similar to the FASTA standard, \pymbe supports other non-standard amino acids beyond the FASTA alphabet with custom parameters provided by the user.
Naturally, \pymbe is not restricted to using pre-defined parameters from the repository but the user is free to define custom parameters of the amino acid particles.

Unlike flexible peptides, coarse-grained models of globular proteins typically use protein structures from the Protein Data Bank (PDB), which are then assumed to be fixed throughout the simulation.
The first step when setting up a globular protein using the \pymbe module is to load its topology.
Currently, the easiest way to load a CG model of a globular protein into \pymbe is to provide a VTF file (VTF Trajectory Format\cite{vtf14a}) with the coordinates of the beads and the information about the protein topology.
To help creating such a VTF file, we provide a supporting script in the \pymbe repository, \code{lib/create\_cg\_from\_pdb.py}.
 This script creates a VTF file with the coarse-grained model of a protein from its corresponding PDB file
\begin{minipage}{\linewidth}
\begin{lstlisting}[language=python3,linewidth=\columnwidth,caption={Using the supporting script to create a pre-defined CG model of a globular protein from a PDB file.}]
python3 create_cg_from_pdb.py 
    --filename FILENAME 
    --download_pdb PDB_CODE 
    --model MODEL 
    --chain_id CHAIN_ID 
\end{lstlisting}
\end{minipage}
The script supports two options to read a PDB file of a protein: (1) from a local PDB file using the argument \texttt{-{}-filename} or (2) by downloading the PDB file from the database, using the argument \texttt{-{}-download\_pdb} and providing the PDB code. 
Optionally, the user can select a sub-chain of the protein by providing the argument \texttt{-{}-chain\_id}.

From the PDB coordinates, the script creates a CG model consisting of either one bead per amino acid (\texttt{-{}-model 1beadAA}) or two beads per amino acid (\texttt{-{}-model 2beadAA}), based on the same logic as explained for peptides.
For the two bead model, the script uses the same procedure as described by Torres \etal\cite{torres19a,torres22a}
One backbone bead is placed at the position of the $\alpha$-carbon of the amino acid with a radius equal to the radius of the alpha carbon. 
A second bead is placed at the center of mass of the side chain of the amino acid residue with a radius equal to its radius of gyration.
For the one-bead model, one bead is placed in the center of mass of each amino acid residue.
We chose this particular representation in order to enable validation of the \pymbe module against already published simulation results.
In the future, we plan to make this procedure more general, allowing the user to define a custom coarse-graining approach.

\begin{figure}[tbp]
    \centering
    \includegraphics[width=\columnwidth]{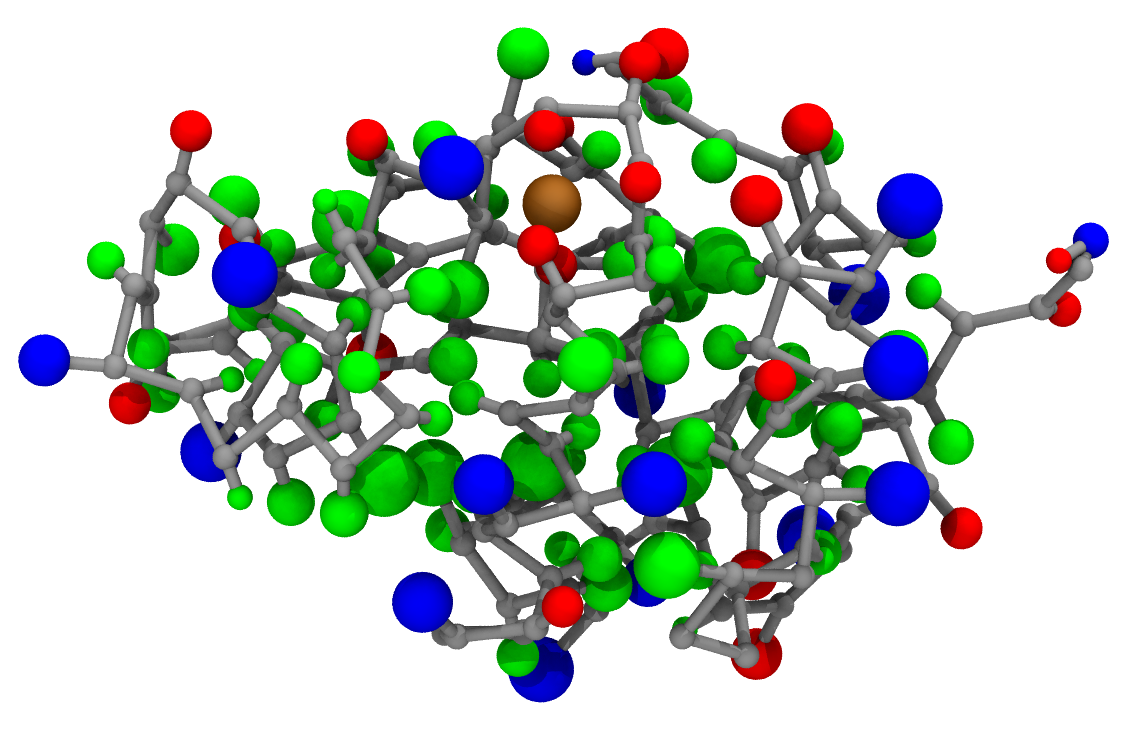}
    \caption{Snapshot of the coarse-grained model of $\alpha$-lactalbumin (PDB code 1F6S \cite{chrysina2000}) using two beads per amino acid. 
    An additional particle is placed in the interior of the protein, representing the $\ce{Ca}^{2+}$ ion caged in the protein structure.
    The particles are colored as follows: acidic bead (red), basic bead  (blue), inert side chain bead (green), inert backbone bead (grey) and $\ce{Ca}^{2+}$ ion (copper).
    The size of the beads has been scaled by an arbitrary value for visualization purposes.
    }
    \label{fig:cg-protein}
\end{figure}

The script creates a VTF file with all coordinates and radii of the beads.
We provide examples of VTF files in the \pymbe repository in the folder \mbox{`\texttt{/parameters/globular\_proteins/}'}. 
To read such files, the \pymbe module provides a function:
\begin{lstlisting}[language=pymbe,linewidth=\columnwidth,,caption={Reading a protein structure from a VTF file.},label={lst:read-topology}]
topology = pmb.read_cg_protein_model(
    filename = protein_filename)
\end{lstlisting}
The function returns a dictionary (\code{topology}) which can be used as an input to another helper function that defines both protein particles and residues in \pymbe:
\begin{lstlisting}[language=pymbe,linewidth=\columnwidth,caption={Defining a protein.},label={lst:define-protein}]
pmb.define_protein( 
    name = "my_protein_1F6S", 
    topology_dict = topology,  
    model = "2beadAA")
\end{lstlisting}
Internally, this function defines one residue and one or two particles for all different amino acids in the protein but it does not define particles and residues for those amino acids which are not present in the protein.
In Fig.~\ref{fig:cg-protein}, we depict an example of a coarse-grained model of $\alpha$-lactalbumin (PDB code 1F6S\cite{chrysina2000}) built using \pymbe, consisting of a rigid object with 2 beads per amino acid (\code{model = "2beadAA"}).

\subsection{\label{sec:methods} Setting up acid-base reactions} 
The \pymbe module supports the automated configuration of two Monte Carlo methods to sample reversible acid-base  reactions: the constant-pH method\cite{reed92a} for the simulation of acid-base equilibria in a single phase and the grand-reaction method,\cite{landsgesell20b, landsgesell20b-err} specifically designed for two-phase systems.
In the following subsections, we give a short introduction to these methods and their usage in \pymbe.
Later, we demonstrate the use of the constant-pH method in our benchmarks for charge-regulating peptides (Section \ref{sec:case_study_peptides}) and globular proteins (Section \ref{sec:case_study_proteins}) in bulk solution.
In Section~\ref{sec:case_study_weak_pe} we showcase the use of the \grxmc method that we use to benchmark weak polyelectrolytes under dialysis.

\begin{figure}
\centering
\includegraphics[width=\columnwidth]{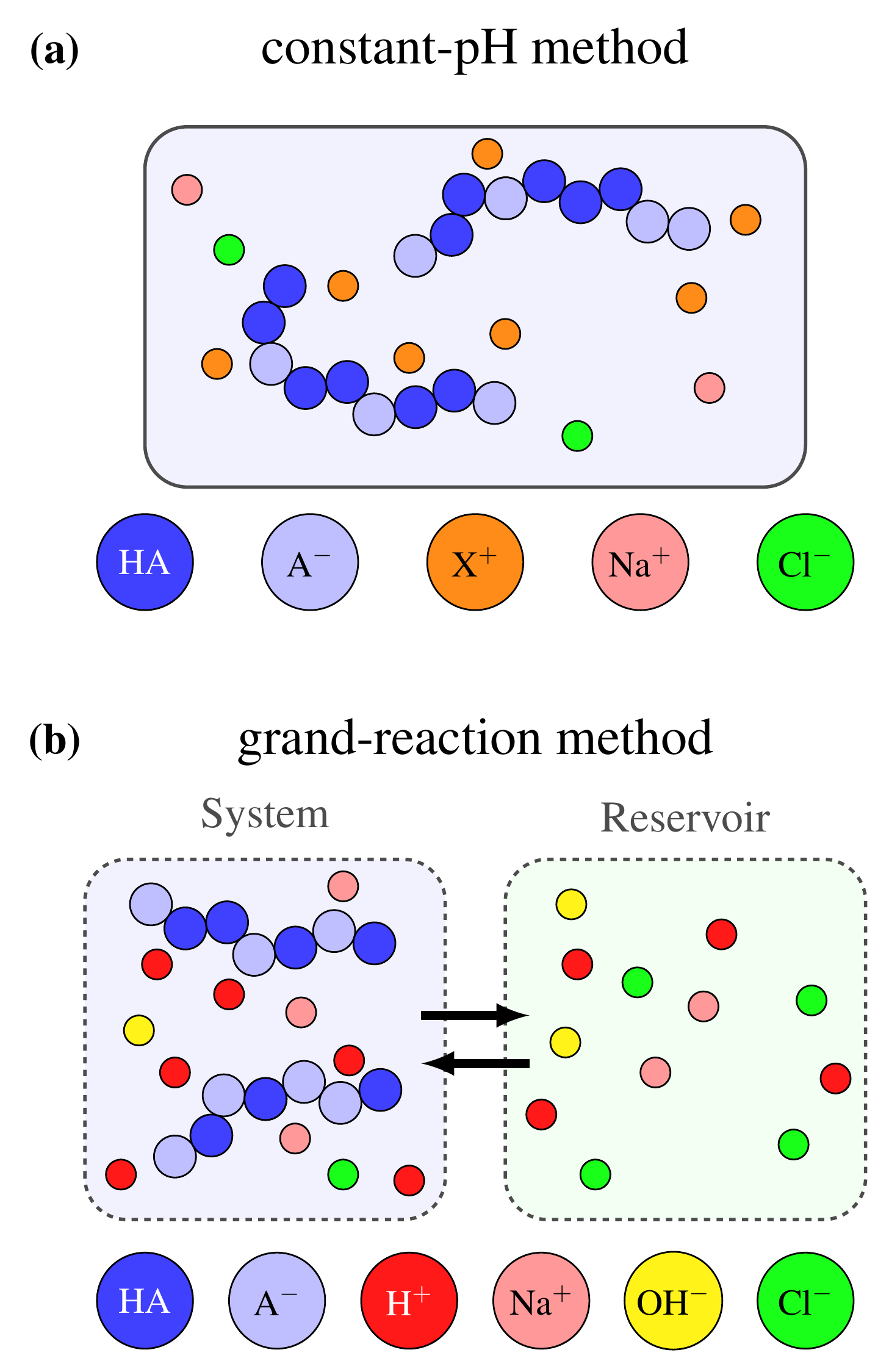}
\caption{\label{fig:schematic_ensembles} Schematic representation of weak polyelectrolyte systems (here represented by weak polyacids with monomers $\ce{HA}$) in different statistical ensembles. (a): In the constant-pH-ensemble (cpH), a single-phase system at a given buffer pH and salt concentration is considered. (b): In the grand-reaction method (G-RxMC), a two-phase system at a fixed reservoir composition is considered. The polyelectrolyte chains cannot leave the "system" phase, while small ions can be partitioned between the "system" and the "reservoir", leading to a Donnan equilibrium.}
\end{figure}

\subsubsection{\label{sec:cpH}The constant-pH method (cpH)}
The constant-pH (cpH) method was originally developed by Reed and Reed\cite{reed92a} to simulate the pH-response of weak polyelectrolyte chains at a given pH in a buffer solution, as shown schematically in Fig. \ref{fig:schematic_ensembles} (a). 
In the cpH method, the pH of the buffer solution is only considered implicitly and it is a direct input parameter of the method.
The method uses Monte Carlo moves to sample different ionization states of weak acid and/or base groups according to the chemical equations
\begin{align}
\ce{HA <=> &A- + X+}\label{eq:acid_equilibria}\\
\ce{HB+ <=> &B + X+},\label{eq:base_equilibria}
\end{align}
with the protonated state of an acid ($\ce{HA}$) and a base ($\ce{HB+}$) and the corresponding deprotonated states $\ce{A-}$ and $\ce{B}$.
\ce{X+}  is a generic neutralizing counterion whose chemical identity could correspond to any monovalent cation (\ce{H+}, \ce{Na+}, \ce{K+}, etc.). 
Since in the cpH method the pH is only considered implicitly, it is decoupled from the activity of the explicit \ce{X+} ions.
This feature of the algorithm limits the applicability of the cpH method to  pH-values where the ionic screening is dominated by the added salt rather than by $\ce{H+}$ or $\ce{OH-}$ ions.\cite{landsgesell17a, landsgesell19a, kosovan23a}
In the cpH method, trial moves consist of changing the ionization state of a charge-regulating particle and randomly inserting/deleting a neutralizing counterion according to \autoref{eq:acid_equilibria} or \autoref{eq:base_equilibria}. These trial moves are accepted with a probability given by\cite{reed92a}
\begin{equation}
P_{\text{cpH}} = \text{min} \left[ 1, \exp\left(-\beta \Delta U + \xi \ln(10) \left( \pH - \pKa \right)\right)  \right]. 
\label{eq:prob_cpH}
\end{equation}
Here, $\beta = \num{1} / k_\text{B}T$ is the inverse thermal energy, $\xi$ the extent of reaction ($\xi=+1$ for a deprotonation and $\xi=-1$ for a protonation) and $\Delta U$ is the change in potential energy between the old state and the proposed new state.
$\pH = \log_{10} (a_{\ce{H}})$ and $\pKa = \log_{10} (K_{\ce{a}})$, where $a_{\ce{H}}$ is the activity of proton and $K_{\ce{a}}$ is the acid dissociation equilibrium constant.
One of the outputs of cpH-simulations is the average degree of protonation of each acidic and basic group.
Notably, the above implementation produces results which are systematically shifted on the pH scale with respect from the correct result, as explained in Refs.\cite{labbez07b,kosovan23a}
In many situations, this deviation is small enough to be neglected.
Nevertheless, it should be considered when presenting results obtained using the method.

The \pymbe module automates the setup of the cpH method in \espresso for all particles that the user defined as having either an \code{"acidic"} or a \code{"basic"} \code{acidity}:
\begin{lstlisting}[language=pymbe,linewidth=\columnwidth,caption={Setting up the cpH method.},breaklines=true]
cpH, labels = pmb.setup_cpH(
    counter_ion = cation_name,
    constant_pH = pH)
\end{lstlisting}
As inputs, one has to provide the \code{name} identifier of the generic neutralizing counterion $\ce{X+}$ (\code{counter_ion}) and the pH at which the system is simulated (\code{constant_pH}). 
The $\pKa$-values of each type of acidic and basic particle must have been previously defined for each particle, as explained in Section~\ref{sec:def_part}.
It returns an instance of the \code{reaction_methods.ConstantpHEnsemble} object of the \espresso library (\code{cpH}) and a list containing the \code{name} of the particles for which reactions have been set up (\code{labels}).

\subsubsection{\label{sec:GRxMC}The Grand-Reaction method (\grxmc)}
In contrast to the cpH method, the \grxmc method developed by Landsgesell \etal\cite{landsgesell20b, landsgesell20b-err} was conceived to study the behavior of two-phase systems, as shown schematically in Fig. \ref{fig:schematic_ensembles} (b).
In these systems, the charge-regulating macromolecules are confined to one phase, which we term the "system". This confinement can be due to covalent crosslinking in the case of hydrogels,\cite{landsgesell22a,beyer22a}
due to a semi-permeable membrane in the case of dialysis,\cite{landsgesell20b} or due to spontaneous phase separation in the case of polyelectrolyte complexes.\cite{stano23a,stano21a}
While the macromolecules are confined to the system phase, small ions ($\ce{H+}$, $\ce{Na+}$, $\ce{OH-}$, $\ce{Cl-}$) can freely partition between the system and the other phase.
This other phase is termed the "reservoir" and it contains only small ions and other solutes which can be exchanged with the system.
The macroscopic constraint that each of the phases has to be electroneutral leads to an uneven partitioning of small ions between the system and the reservoir, the so-called Donnan equilibrium.\cite{landsgesell20b}

The \grxmc method was designed to mimic a setting where the reservoir is much larger than the system and thus the reservoir properties are not affected by the partitioning of ions.
This allows one to to study the properties of the system at a given fixed composition of the reservoir.
By construction, the reservoir composition in the grand-reaction method is specified using linear combinations of the chemical potentials of the exchangeable species (ions).
The coupling of the system to the reservoir can be formally represented by virtual chemical reactions:\cite{landsgesell20b}
\begin{align}
\emptyset \ce{<=> &H+ + OH-}\label{eq:insertion_water}\\
\emptyset \ce{<=> &Na+ + Cl-}\label{eq:insertion_nacl}\\
\emptyset \ce{<=> &Na+ + OH-}\label{eq:insertion_naoh}\\
\emptyset \ce{<=> &H+ + Cl-}\label{eq:insertion_hcl},
\end{align}
where $\emptyset$ denotes an empty set.
The above reactions effectively result in the insertion or deletion of the respective ion pairs in the system, conserving the overall electroneutrality.
Furthermore, the rules for the acceptance probability of these insertions ensure that the chemical potential of each ion pair is the same as in the reservoir.\cite{landsgesell20b}
It can be shown that these insertions are equivalent to a grand-canonical simulation.
If only the exchange of salt ions with the reservoir is desired, the function \code{pmb.setup_gcmc} allows to set up a pure grand-canonical Monte Carlo (GCMC) simulation.
However, if there are also acid-base reactions taking place in the system,
\begin{align}
\ce{HA <=> &A- + H+}\label{eq:acid_equilibria_grxmc}\\
\ce{HB+ <=> &B + H+},\label{eq:base_equilibria_grxmc}
\end{align}
then one needs to set up the G-RxMC method instead. The acceptance probability for Eqs. \ref{eq:insertion_water}-\ref{eq:base_equilibria_grxmc} are given by\cite{smith94c, johnson94a,landsgesell20b}
\begin{equation}
P_{\text{RxMC}} = \text{min} \left[ 1, \exp\left(-\beta \Delta U \right) K_\text{c}^\xi \left( V N_\text{A} \right)^{\bar{\nu}\xi} \prod_i \frac{N_i!}{(N_i+\nu_i \xi)!} \right].
\label{eq:prob_RxMC}
\end{equation}
In this equation, $K_\text{c}$ is the concentration-based equilibrium constant of the considered reaction, $V$ denotes the box volume, $N_\text{A}$ is the Avogadro constant, $N_i$ the number of particles of type $i$, $\nu_i$ is their stoichiometric coefficient and $\bar{\nu} = \sum_i \nu_i$.

Notably, the numerical value of $K_\text{c}$ depends on the chosen reference concentration, which in the system of reduced units of \espresso corresponds to \code{1/unit_length**3} (cf. Code snippet \ref{lst:custom-units}).
Therefore, the user needs to convert the equilibrium constants from the literature, based on the reference concentration $c^{\ominus} = \qty{1}{mol/kg}\approx\qty{1}{mol/L}$, to the system of reduced units used in \espresso.\cite{landsgesell19a}
This conversion often results in errors which can be avoided by the tools provided in the \pymbe module.
Furthermore, there is a complicated relation between the desired reservoir composition (pH-value and salt concentration) and the equilibrium constants required as inputs, as explained in Appendix \ref{sec:reservoir_composition_grxmc}.
It can be determined numerically either by running a set of auxiliary simulations of the reservoir\cite{landsgesell20b, beyer23b} or by using an approximate theoretical model for the excess chemical potential of a salt solution, such as the Debye--Hückel limiting law.
Only then, the user can run G-RxMC simulations at a given salt concentration and a given pH of the reservoir.
\pymbe currently only supports setting up a reservoir containing a monovalent salt. 
In principle, the G-RxMC method can also accommodate multivalent salts, but the setup becomes more complicated in this case.\cite{beyer23b}

The \pymbe module provides a convenient and user-friendly way of specifying the reservoir composition and the corresponding equilibrium constants.
In \pymbe, the user specifies the desired pH in the reservoir \mbox{(\code{pH_res})} and the concentration of added monovalent salt (e.g. $\ce{NaCl}$) in the reservoir (\code{c_salt_red}) in the following way:
\begin{minipage}{\linewidth}
\begin{lstlisting}[language=pymbe,linewidth=\columnwidth,caption={Setting up the G-RxMC method.},breaklines=true]
grxmc, labels, I_res = pmb.setup_grxmc(
    pH_res = 2, 
    c_salt_res = salt_conc_res, 
    proton_name = "H",
    hydroxide_name = "OH",
    salt_cation_name = "Na",
    salt_anion_name = "Cl"
    activity_coefficient = f_act_coeff)
\end{lstlisting}
\end{minipage}
In addition to the concentration of added salt and pH in the reservoir, this function takes as required arguments the names of the various small ions (\code{proton_name},  etc.).
The argument \code{activity_coefficient} is a function that calculates the activity coefficient of an ion pair in a solution of ions as a function of the ionic strength.
For example, using the Debye--Hückel limiting law for the activity coefficient in an aqueous solution at standard conditions,\cite{atkins23a} a possible implementation of the function \code{activity_coefficient} would look like this:
\begin{lstlisting}[language=pymbe,linewidth=\columnwidth,caption= {Calculating the activity coefficient.},breaklines=true]
def f_act_coeff(I):
    from numpy import exp, log, sqrt
    I_mag = I.to("mol/L").magnitude
    A = 0.509
    act = exp(-2*log(10)*A*sqrt(I_mag))
    return act
\end{lstlisting}
Alternatively, one can use a more sophisticated model such as the Davies equation\cite{atkins23a} or data of the excess chemical potential from simulations, e.g. obtained using the method of Widom particle insertion.\cite{widom63a}
Given the above inputs, \pymbe internally solves a coupled system of nonlinear equations, relating the required equilibrium constants with \code{pH_res}, \code{c_salt_res} and \code{activity_coefficient}, until a self-consistent solution is reached (more details are provided in Appendix \ref{sec:reservoir_composition_grxmc}).
When solving these equations, \pymbe assumes that the solution contains only monovalent salt and that the pH was adjusted to the desired value by adding a strong base (NaOH) or a strong acid (HCl) to the salt solution.
Eventually, it uses the determined equilibrium constants to set up the full set of reactions required for the G-RxMC simulation.
Overall, the function \code{setup_grxmc} greatly simplifies the setup of G-RxMC simulations, helping the user to automate multiple non-trivial steps that are susceptible to errors.

Instead of distinguishing between different ion types of the same charge (which differ only in their label but not with regards to their interactions in the coarse-grained representation), one can also formulate the \grxmc method in terms of unified ion types.\cite{curk22a}
To set up the G-RxMC method using the implementation in terms of unified ion types, the function \code{setup_grxmc_unified} is employed in fashion analogous to \code{setup_grxmc}.

\subsection{\label{sec:interactions}Setting up interactions in \espresso with \pymbe}
Before creating all the molecules in \espresso, the user needs to provide some additional information about the desired interactions.
First, parameters of the bonding potentials need to be provided using the following function:
\begin{lstlisting}[language=pymbe,linewidth=\columnwidth,caption={Defining bonds.},breaklines=true]
# Define the parameters of the bond
prm={"r_o": 0.5*pmb.units.nm,
     "k": 0.4*pmb.units.N/pmb.units.m}

pmb.define_bond(
    bond_parameters = prm,
    bond_type= "harmonic"
    particle_pairs = [["I","I"],
                     ["I","A"]])
\end{lstlisting}
Here, \code{bond_type} is a string identifying the bond type.
Currently, \pymbe supports the harmonic potential (\code{bond_type =  "harmonic"}) and the Finite Extension Non-linear Elastic potential (\code{bond_type = "FENE"}) as model interactions for  bonding the particles.\cite{weik19a,weeber24a}
\code{bond_parameters} is a dictionary with the values and the keys of the parameters for the potential energy of the bond.
For consistency, the variable names of the parameters of each type of bond are the same as the ones in \espresso.
\code{particle_pairs} is a list with a set of pairs of particle names which identifies types of particles to be bonded by \pymbe with the defined bond type.
All bonds defined in \pymbe are stored in the \pymbe dataframe, as explained in Appendix \ref{sec:appendix_pymbe_df}.

The user can add all bonds defined in the \pymbe dataframe in the \espresso system using the helper function
\begin{lstlisting}[language=pymbe,linewidth=\columnwidth,caption={Adding bonds to ESPResSo.},breaklines=true]
pmb.add_bonds_to_espresso(
    espresso_system = es_system)
\end{lstlisting}
Here, \code{espresso_system} is an instance of an \espresso system from the \espresso library.

The \pymbe module has another helper function that  automatically sets up Lennard-Jones (LJ) interactions between the particles using the following pairwise potential
\begin{widetext}
\begin{equation}
\begin{split}\label{eq:lj}
  V_{ij}(r) =&
    \begin{cases}
      \infty & \text{for}\ r < r_\mathrm{off}\\
      4 \epsilon \left[ \left(\displaystyle\frac{\sigma}{r-r_\mathrm{off}}\right)^{12}
      - \left(\displaystyle\frac{\sigma}{r-r_\mathrm{off}}\right)^6\right] - C
      & \text{for}\ r_\mathrm{off} < r < r_\mathrm{cut}+r_\mathrm{off}\\
      0 & \text{for}\ r > r_\mathrm{cut}+r_\mathrm{off}
    \end{cases},\end{split}
\end{equation}
\label{eq:LJ}
\end{widetext}
where $r$ is the distance between the interacting particles $i$ and $j$, $\sigma$ determines the effective range of the repulsion, $\epsilon$ is the depth of the attractive well, $r_\mathrm{cut}$ is the distance at which the potential is cut off to 0 and $r_\mathrm{off}$ is an offset that shifts the values of the potential. 
The constant $C = 4\epsilon( (\sigma / r_\mathrm{cut})^{12} - (\sigma / r_\mathrm{cut})^{6} )$ is added to make the potential continuous at $r_\mathrm{cut}$.
The augmented LJ potential presented in Eq. \ref{eq:LJ} reduces to the classical LJ potential with $r_\mathrm{off} = 0$ and $r_\mathrm{cut} \rightarrow \infty$.
The user can set up the LJ potential between all particles which have been previously defined in \pymbe using the helper function
\begin{lstlisting}[language=pymbe,linewidth=\columnwidth,caption={Setting up the Lennard-Jones interactions.},breaklines=true]
pmb.setup_lj_interactions(
    espresso_system = es_system)
\end{lstlisting}
The values of the LJ parameters $\sigma$, $\epsilon$, $r_\mathrm{cut}$ and $r_\mathrm{off}$ for each particle pair are calculated according to the Lorentz--Berthelot combining rules\cite{lorentz81a,berthelot98a} using the 
corresponding \mbox{\code{sigma},} \code{epsilon}, \code{cutoff} and \code{offset} parameters provided by the user when defining each particle type in \pymbe (cf. Code snippet \ref{lst:define-particle}).
By default, \pymbe sets $r_{\mathrm{cut}}=\sqrt[6]{2}\sigma$ and $r_{\textrm{off}} = 0$, corresponding to a purely repulsive LJ potential, also known as the Weeks-Chandler-Andersen potential.
If the user defines a particle with \code{sigma = 0}, \pymbe will not set up LJ interaction between that particle type and any other particle, corresponding to the case of an ideally behaving particle. 
If the user has not defined either the value of \code{sigma} or \code{epsilon} of a particle type (cf. Code snippet \ref{lst:define-particle}), \pymbe will not set up any LJ interaction between particles of that type and the rest of particles.

\subsection{\label{sec:setup}Setting up an \espresso system with \pymbe}
After defining particles, residues, molecules, chemical reactions and interactions, everything is ready to create the particle objects in \espresso.
However, to make it all work, the above steps have to be executed in a particular order.
First, one should define all particles, residues and molecules:
\begin{lstlisting}[language=pymbe,linewidth=\columnwidth,caption={Defining the properties of the different types of particles, residues and molecules in the model.}]
pmb.define_particle(...)
pmb.define_residue(...)
pmb.define_molecule(...)
\end{lstlisting}
In the next step, one defines the interactions and chemical reactions, which requires that the corresponding particles and residues have been previously defined in \pymbe:
\begin{minipage}{\linewidth}
\begin{lstlisting}[language=pymbe,linewidth=\columnwidth,caption={Defining the interactions in the model and setting up chemical reactions.}]
pmb.define_bond(...)
pmb.add_bonds_to_espresso(...)
pmb.setup_lj_interactions(...)
pmb.setup_cpH(...) # for cpH
#pmb.setup_grxmc(...) for G-RxMC
\end{lstlisting}
\end{minipage}
Note that if one defines new particles after setting up LJ interactions, then these new particles will not be assigned any interactions, which will probably lead to an undesired behavior.
In contrast, activating electrostatic interactions should be done only after creating all objects in \espresso and relaxing the initial simulation setup.

After completing the above steps, \pymbe is ready to create these objects in the \espresso system.
For example, if \code{"alternating_polymer"} has been defined (cf. Code snippet~\ref{lst:list-molecules}), it can be created using the following helper function:
\begin{lstlisting}[language=pymbe,linewidth=\columnwidth,caption={Creating a pyMBE object in ESPResSo.},breaklines=true]
pmb.create_pmb_object(
    name = "alternating_polymer",
    espresso_system = es_system,
    number_of_objects = 2)
\end{lstlisting}
In the above Code snippet, \code{name} is the \pymbe string identifier of the previously defined \pymbe object which will be created in \espresso.
In practice, it means that the corresponding particles are created in \espresso system and bonds are assigned to the specific particle pairs.
The parameter \code{number_of_objects} is the number of individual objects to be created.
Currently, the above function can create the following types of \pymbe objects: particles, residues, molecules and peptides.
By default, these objects will be created in the \espresso system at a random position in the simulation box.
By providing an optional argument \code{position}, the user can specify the position of the first particle of a given object.
For simplicity, the initial state of flexible molecules is fully stretched.
For acidic and basic particles, whose charges fluctuate during the simulation, the initial charge state is set to match that of the protonated form, \ie $q=0$ for an acid and $q=+1e$ for a base.
In principle, the choice of the initial state is arbitrary and its impact on the simulated system should vanish during equilibration of the system.
When creating these objects, \pymbe assigns a specific identifier to each particle, residue and molecule created in \espresso system.
This identifier can be accessed via the \pymbe dataframe, as described in the next section.

For the purpose of creating rigid models of globular proteins in \espresso, the \pymbe module uses a different function.
Once a protein topology has been loaded to \pymbe, (cf. Code snippet~\ref{lst:define-protein}), it can be created in the \espresso system as follows:
\begin{lstlisting}[language=pymbe,linewidth=\columnwidth,caption={Creating a protein in ESPResSo.},breaklines=true]
pmb.create_protein(
    name = "my_protein_1F6S", 
    number_of_proteins = 1, 
    espresso_system = es_system,
    topology_dict = topology)
\end{lstlisting}
Here, \code{name} is the identifier of the protein object previously defined in \pymbe, \code{espresso_system} is an instance of an \espresso system from the \espresso library and \code{number_of_proteins} is the number of proteins of that type to be created in \code{es_system}. 
By default, the proteins are created at random positions.
Similar to the previous case, \pymbe assigns a unique identifier to each created particle, residue and protein, and this identifier can be accessed via the \pymbe dataframe, as described in the next section.

When creating proteins in \espresso, \pymbe uses the rigid object framework of \espresso to ensure that the protein as a whole remains rigid, \ie it does not deform.
Additionally, \pymbe fixes the position of all particles belonging to the proteins.
This setup is convenient for simulations with a single protein in the simulation box, for which the module features the option to center the protein in the simulation box: 
\begin{lstlisting}[language=pymbe,linewidth=\columnwidth,caption={Centering a molecule.},breaklines=true]
pmb.center_molecule_in_simulation_box(
    molecule_id = protein_id, 
    espresso_system = es_system)  
\end{lstlisting}
Here, \code{molecule_id} is the numeric identifier within the \pymbe dataframe of the protein to be centered in the simulation box. 
Although this function was designed for this particular application, its implementation is general and it can be used with any molecule defined using the \pymbe module.
To set up simulations having multiple proteins in the simulation box, \pymbe permits to enable the motion of the rigid object as a whole through the simulation box
\begin{lstlisting}[language=pymbe,linewidth=\columnwidth,caption={Enabling motion.},breaklines=true]
pmb.enable_motion_of_rigid_object(
    espresso_system = es_system, 
    name = protein_name)
\end{lstlisting}
where \code{name} is the identifier of the type of proteins whose motion should be enabled.

\subsection{\label{sec:pymbe_df} Accessing the information stored in \pymbe} 
The \pymbe module internally uses a Pandas dataframe\cite{mckinney2010a,pandas2023} to store all information about the particles, residues, molecules, bonds and interactions defined by the user.
It also stores all information about additional parameters which have been defined automatically by calling various helper functions.
The dataframe is stored is stored as an attribute of the \pymbe library and it can be accessed as follows:
\begin{lstlisting}[language=pymbe,linewidth=\columnwidth,caption={Example of how to access the pyMBE dataframe to print it.},breaklines=true]
print(pmb.df)
\end{lstlisting}
The \pymbe dataframe can be operated on as a standard Pandas dataframe.
It allows the user to easily access all parameters defined in the \pymbe module. 
To facilitate this access, the user can filter the dataframe by specifying the \pymbe object of interest in the argument \code{pmb_type}: 
\begin{lstlisting}[language=pymbe,linewidth=\columnwidth,caption={Example of how to filter the pyMBE dataframe by particle.},breaklines=true]
pmb.filter_df(pmb_type = "particle")
\end{lstlisting}
The argument \code{pmb_type} supports \pymbe objects of the following types: \code{"particle"}, \code{"residue"}, \code{"molecule"} and \code{"bond"}.
In Table \ref{tab:df_example_par}, we show an example of the output of this function when filtering by \code{pmb_type = "particle"}.
In this example, each column corresponds to a parameter of a \pymbe particle object and each row to a different particle in the system. 
Additional examples of subsets of the \pymbe dataframe can be found in Appendix \ref{sec:appendix_pymbe_df}.

\begin{table*}[]
\caption{\label{tab:df_example_par} 
Example of a subset of the \pymbe dataframe, filtered by \code{pmb_type = "particle"}. 
}
\footnotesize
\resizebox{\textwidth}{!}{%
\begin{tabular}{@{}cccccccccccccccc@{}}
\toprule\addlinespace[0.4em]
\multirow{2}{*}{\textbf{name}} & \multirow{2}{*}{\textbf{pmb\_type}} & \multirow{2}{*}{\textbf{particle\_id}} & \multirow{2}{*}{\textbf{residue\_id}} & \multirow{2}{*}{\textbf{molecule\_id}} & \multirow{2}{*}{\textbf{acidity}} & \multirow{2}{*}{\textbf{pKa}} & \multirow{2}{*}{\textbf{sigma}} & \multirow{2}{*}{\textbf{epsilon}} & \multicolumn{3}{c}{\textbf{state\_one}} & \multicolumn{3}{c}{\textbf{state\_two}} & \multirow{2}{*}{$\cdots$}  \\\addlinespace[0.2em]
        &          &   &   &   &        &     &        &               & \textbf{label}    & \textbf{es\_type} & \textbf{z} & \textbf{label}   & \textbf{es\_type} & \textbf{z} & \\ \addlinespace[0.4em]\midrule\addlinespace[0.4em]
I & particle & 0 & 0 & 0 & NaN  & NaN & 0.3 nm & 1 reduced\_energy & I  & 0        & 0      & NaN     & NaN      & NaN & $\cdots$   \\ \addlinespace[0.8em]
A & particle & 1 & 0 & 0 & acidic & 4.0 & 0.4 nm & 1 reduced\_energy & AH & 1        & 0      & A    
   & 2        & -1  & $\cdots$   \\ \addlinespace[0.8em]
B & particle & 2 & 0 & 0 & basic  & 9.0 & 0.5 nm & 1 reduced\_energy & BH & 3        & +1     & B       & 4        & 0 & $\cdots$ \\
$\vdots$ & $\vdots$ & $\vdots$ & $\vdots$ & $\vdots$ & $\vdots$  & $\vdots$  & $\vdots$ & $\vdots$ & $\vdots$ & $\vdots$ & $\vdots$ & $\vdots$ & $\vdots$ & $\vdots$ & $\ddots$\\\bottomrule
\end{tabular}%
}
\end{table*}

Internally, the \pymbe module assigns a numerical identifier to each individual particle (\code{particle_id}), residue (\code{residue_id}) or molecule (\code{molecule_id}) object that \pymbe creates in the \espresso system. 
These identifiers are stored in the \pymbe dataframe in which \pymbe creates one row per each individual object.
They permit to map which specific particles belong to specific residues and molecules, allowing to trace back the exact topology of the system.
The identifiers \code{state_one} and \code{state_two} store information about the possible ionization states of the particles. 
Each state has a secondary header with the following indexes: \code{label}, \code{es_type} and \code{charge}. 
The string-like identifier \code{label} labels a particular ionization state of the particle.
The numerical identifier \code{es_type} stores the particle type used in \espresso for that particular type of \pymbe particle.
The parameter \code{charge} stores the valency of a particle in a given state.
Inert particles are defined with only one state (\code{state_one}) whose label matches with its \code{name} used by the \pymbe module.
For acidic and basic particles, \code{state_one} corresponds to the protonated state and \code{state_two} to the deprotonated state. 
The \pymbe module labels \code{state_one} with the \code{name} of the particle with an extra "H" character at the end, while \code{state_two} is labelled using the same label as in the \code{name} of the particle.  
Additionally, \pymbe assigns the charge of each state using the \code{acidity} of the particle. 
For an acidic particle, \code{charge} is set to 0 in \code{state_one} (corresponding to a protonated monoprotic acid) and \code{charge} is set to to $-1e$ in \code{state_two} (corresponding to a deprotonated monoprotic acid). 
For a basic particle, \code{charge} is set to $+1e$ in \code{state_one} (corresponding to a protonated monoprotic base) and \code{charge} is set to $0$ in \code{state_two} (corresponding to a deprotonated monoprotic base).

The user can store the \pymbe dataframe in a local file 
\begin{lstlisting}[language=pymbe,linewidth=\columnwidth,caption={Locally storing the pyMBE dataframe.},breaklines=true]
pmb.write_pmb_df(filename)
\end{lstlisting}
This function uses the Pandas native function \code{pandas_df.to_csv(filename)} to save the \pymbe dataframe into a file in CSV format.
Due to the multi-index organization of the \pymbe dataframe, the user cannot directly use the Pandas native function to load files in CSV files. 
However, the \pymbe module provides a helper function to read the information from a \pymbe dataframe stored in a local CSV file 
\begin{lstlisting}[language=pymbe,linewidth=\columnwidth,caption={Reading an existing pyMBE dataframe.},breaklines=true]
pmb.read_pmb_df(filename = "df.csv")
\end{lstlisting}
where \code{filename} is the path to the local CSV file with the \pymbe dataframe.

\section{Benchmarks}
In this section, we show a set of benchmark results, where we used \pymbe to reproduce the results from some of our previous publications.
Some of these previous results were obtained using \espresso, while others were obtained using different software packages.
Nonetheless, their common feature is that they all calculated the net charge of various molecules containing weakly acidic or basic groups.
Before describing the results, which we used as benchmarks, we describe some additional helper functions, provided by \pymbe to facilitate the calculation of acid-base properties and net charge of molecules in complex simulation setups.

\subsection{Helper functions for analyzing acid-base properties}
As a reference to evaluate the effect of interactions, we use the Henderson--Hasselbalch (HH) equation, which is the analytical solution of the acid-base chemical equilibrium (Eqs. \ref{eq:acid_equilibria} and \ref{eq:base_equilibria}) under ideal conditions, \ie in the absence of interactions. 
Under this assumption, the degree of ionization, $\alpha_i$, of a monoprotic acid or base $i$ is  
\begin{equation}
\alpha_i \ideq \frac{1}{1 + 10^{z_i(\pKai - \pH)}},
\label{eq:alpha_HH}
\end{equation}
where $\Kai=10^{-\pKai}$ is its the acidity constant and $\pH = -\log a_{\ce{H+}}$ is defined in terms of the proton activity $a_{\ce{H+}}$.
The charge number of the group $i$ in the ionized state is denoted by $z_i$, \ie $z_i = +1$ for a base and $z_i = -1$ for an acid. 
The average net charge $Q$ of a macromolecule with $N$ ionizable groups can be calculated by summing over the average degrees of ionization of all its ionizable groups: 
\begin{equation}
Q = \sum^N_{i=1}e z_i \alpha_i.
\label{eq:net_charge}
\end{equation} 
The \pymbe module provides a function that calculates the ideal net charge of any molecule defined with a given \code{molecule_name}:
\begin{lstlisting}[language=pymbe,linewidth=\columnwidth,caption={Calculating the ideal net charge of a molecule.},breaklines=true]
Z_HH = pmb.calculate_HH(
    molecule_name = "molecule_name", 
    pH_list = pH_range)
\end{lstlisting}
Here, \code{pH_list} is the list of pH-values at which the Henderson--Hasselbalch equation is evaluated.
This function is particularly handy for molecules with many different acidic and basic groups, for example peptides and proteins.
Comparison of simulation results with the above ideal theory serves as an important metric for quantifying the effects of interactions on the ionization behavior of macromolecules.

The deviations from the ideal theory, observed in simulations of isolated macromolecules in bulk solution, are predominantly caused by electrostatic interactions between various ionizable groups within the molecule, which has been termed the "polyelectrolyte effect".\cite{landsgesell20b}
In two-phase systems in which macromolecules are confined in one phase (\emph{cf}. Fig. \ref{fig:schematic_ensembles}b), there is an additional contribution to deviations from the ideal behavior, which has been termed the "Donnan effect".\cite{landsgesell20b}
This name refers to the Donnan potential $\psi^\mathrm{Don}$, which is the electrostatic potential difference between the two phases due to the uneven partitioning of the small ions.
The Donnan potential leads to a difference in pH between the system ($\text{pH}^\text{sys}$) and the reservoir ($\text{pH}^\text{res}$),\cite{landsgesell20b, landsgesell22a}
\begin{align}
	\text{pH}^\text{sys} - \text{pH}^\text{res} = \frac{\beta e\psi^\text{Don}}{\ln 10}.
\label{eq:Donnan_effect_pH}
\end{align}
Both Donnan effect and polyelectrolyte effect contribute to the shift of the ionization curves in two-phase systems, whereas only the polyelectrolyte effect is present in single-phase systems (bulk solutions).

The \pymbe module provides a function \code{calculate_HH_Donnan} which allows the user to calculate the charge of the molecules including the Donnan effect in the ideal gas limit, \ie assuming that the acid-base equilibrium can be calculated using the Henderson--Hasselbalch equation:
\begin{lstlisting}[language=pymbe,linewidth=\columnwidth,caption={Calculating the ideal net charge accounting for the Donnan effect.},breaklines=true,label={lst:HH_Don}]
c_peptides = {
    "pep_1": pmb.Quantity(10,"mmol/L"),
    "pep_2": pmb.Quantity(5,"mmol/L")}

Donnan_dict = pmb.calculate_HH_Donnan(
    c_salt  = salt_conc_res, 
    c_macro = c_peptides,
    pH_list = pH_range)
\end{lstlisting}
In contrast to the function \code{calculate_HH}, one needs to provide a dictionary \code{c_macro} containing the concentration \emph{all} charged molecules in the system as well as the salt concentration in the reservoir \code{c_salt}.
For example, the setup in Code snippet \ref{lst:HH_Don} corresponds to a system with two types of peptides: \code{pep_1} and \code{pep_2}.
This additional information is required to correctly calculate the Donnan partitioning, which depends on the ionic strength of the reservoir and the concentration of charges in the system that cannot be exchanged with the reservoir.
The performed calculation consists of solving a system of coupled non-linear equations involving the degrees of ionization according to the the Henderson--Hasselbalch equations for the various ionizable particles in the system and partition coefficients of small ions according to the ideal Donnan theory (See Appendix \ref{sec:HH_coupled_Donnan}).
The function returns a dictionary containing the calculated charges of the various species, a list of the pH-values \emph{in} the system $\text{pH}^\text{sys}$ and a list of the partition coefficients of monovalent cations.

Finally, \pymbe provides a function \code{calculate\_net\_charge}, which enables the calculation of the instantaneous net charge of a given molecule type, existing in the simulation box of \espresso.
If more molecules of such type exist in the simulation box, then net charge is averaged over all molecules of the same type:
\begin{lstlisting}[language=pymbe,linewidth=\columnwidth,caption={Calculating the net charge of a molecule.},breaklines=true]
charge_dict = pmb.calculate_net_charge(
    espresso_system = es_system,
    molecule_name = "molecule_name")
\end{lstlisting}
The returned dictionary contains the mean net charge per molecule, as well as a dictionary containing the mean net charge of the individual residues.
All three helper functions described above can be conveniently used in post-processing of the benchmark results described in the following subsections.

\subsection{Peptides in bulk solution}
\label{sec:case_study_peptides}
As a benchmark of the \pymbe module, we calculated the net charge of various peptides as a function of pH.
The first two peptides, $\mathrm{Lys}_5\mathrm{Asp}_5$ and $\mathrm{Glu}_5\mathrm{His}_5$, each contain a block of five acidic and five basic amino acids. 
These artificial amino acid sequences have been custom synthesized and used by Lunkad \etal\cite{lunkad21a} to validate the net charge predicted by their coarse-grained model against experimental measurements.
These peptide sequences have been selected to exhibit a rather simple response to a change of the pH.
The amino acids in $\mathrm{Lys}_5\mathrm{Asp}_5$ have a rather big difference in $\pKa$ values.
Therefore, the corresponding ionization curve contains two distinct inflection points.
In contrast, the amino acids in $\mathrm{Glu}_5\mathrm{His}_5$ have a much smaller difference in $\pKa$ values, resulting in an almost linear decrease of net charge in a broad range of pH values.
The model used by Lunkad \etal\cite{lunkad21a} consisted of two beads per amino acid. 
Their results were obtained using the constant pH (cpH) ensemble implemented in \espresso.

\begin{figure}[t]
\centering
\includegraphics[width=\columnwidth]{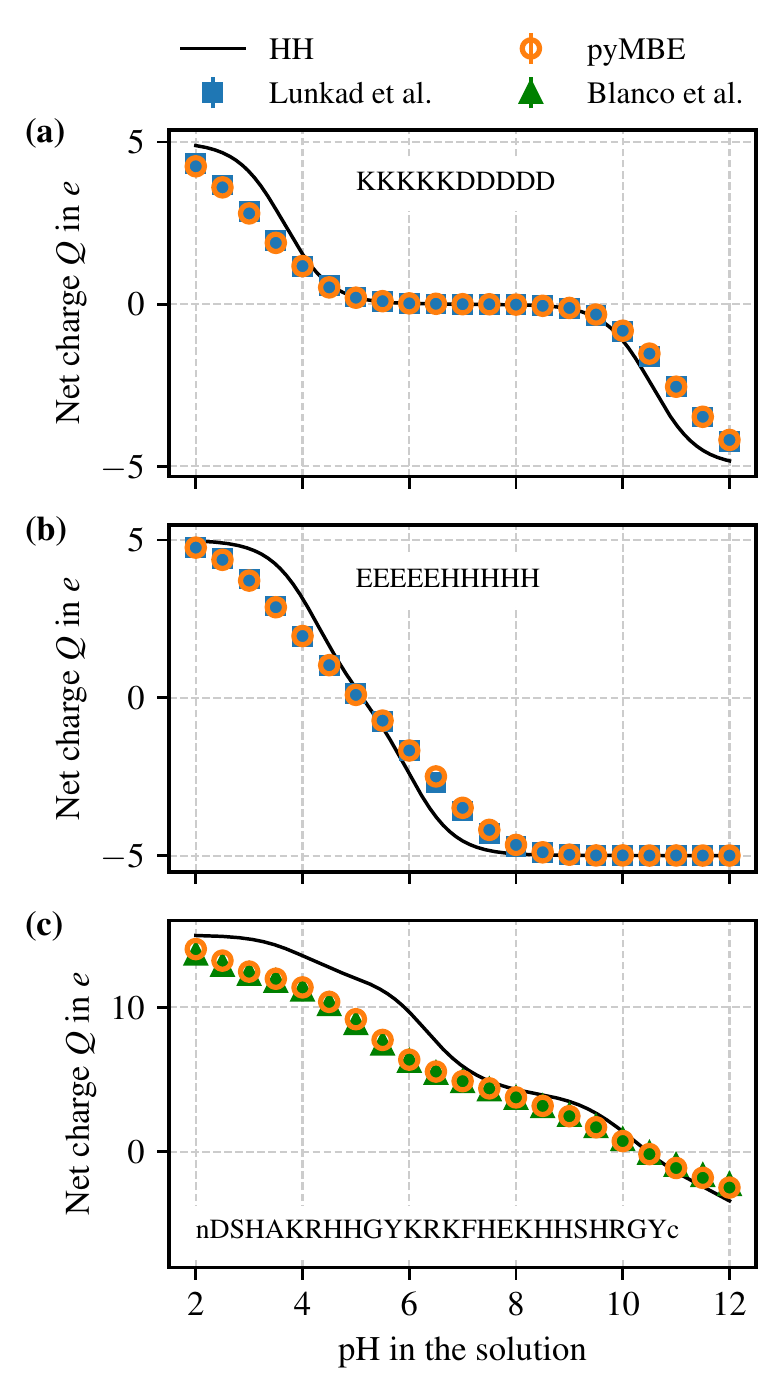}
\caption{\label{fig:peptide_tests} Net charge $Q$ of various peptides as a function of the pH. 
Panels (a) and (b): titration curve of the synthetic peptides $\mathrm{Lys}_5\mathrm{Asp}_5$ (a) and $\mathrm{Glu}_5\mathrm{His}_5$ (b) obtained with constant pH Monte Carlo (cpH) simulation  using \espresso by Lunkad \etal (blue markers).\cite{lunkad21a} 
Panel (c): titration curve of histatin-5 computed by cpH simulation using MOLSIM by Blanco \etal (green markers).\cite{blanco21a} 
The data from these references can be easily reproduced using the \pymbe module to set up these systems in \espresso (orange markers).
The  black lines follow the  Henderson--Hasselbalch (HH) equation (Eqs. \ref{eq:alpha_HH} and \ref{eq:net_charge}) using different sets of $\pKa$-values:  the Chemical Rubber Company (CRC) Handbook of Chemistry and Physics\cite{lide1991a} (a and b) and Nozaki and Tanford \cite{nozaki1967a} (c).}
\end{figure}

As explained in Section \ref{sec:pre-defined}, the \pymbe module allows to easily set up two-bead models of peptides in \espresso, simply by specifying the amino acid sequence and selecting the desired set of model parameters. 
In Fig. \ref{fig:peptide_tests} (a) and (Fig. \ref{fig:peptide_tests} (b)) we show that the simulations with the two-bead model  created using \pymbe quantitatively reproduce the results of Lunkad \etal\cite{lunkad21a}

The third peptide is histatin-5, a natural peptide present in human saliva with antifungal and antibacterial properties.\cite{hyltegren2016a,mackay1984a,puri2014a}
Unlike the previous two, this natural peptide contains a more complex combination of acidic and basic groups.
Specifically, it contains 15 basic groups of 4 different types: N-terminal amine, Lysine, Histidine, Arginine; and 5 acidic groups of 4 different types: C-terminal carboxylic acid, Tyrosine, Aspartic acid and Glutamic acid.
Therefore, the net charge of this protein as a function of pH exhibits a more complex trend.
This peptide was simulated by Blanco \etal\cite{blanco21a} using a one-bead model and constant-pH ensemble simulations performed in MOLSIM. 
In Fig. \ref{fig:peptide_tests}(c) we show that our simulation in the cpH ensemble with the one-bead model created using \pymbe reproduced the original results for histatin-5 from Ref.\cite{blanco21a}

In summary, the above results demonstrate the use of \pymbe for setting up constant-pH simulations of flexible peptides, using the amino acid sequence and pre-defined sets of $\pKa$ values as inputs of the simulations.
Importantly, the use of \pymbe dramatically simplifies setting up of these simulations in \espresso, making it use more user-friendly and less prone to errors.

\subsection{Globular proteins in bulk solution} 
\label{sec:case_study_proteins}
\begin{figure}[t]
    \centering
    \includegraphics[width=\columnwidth]{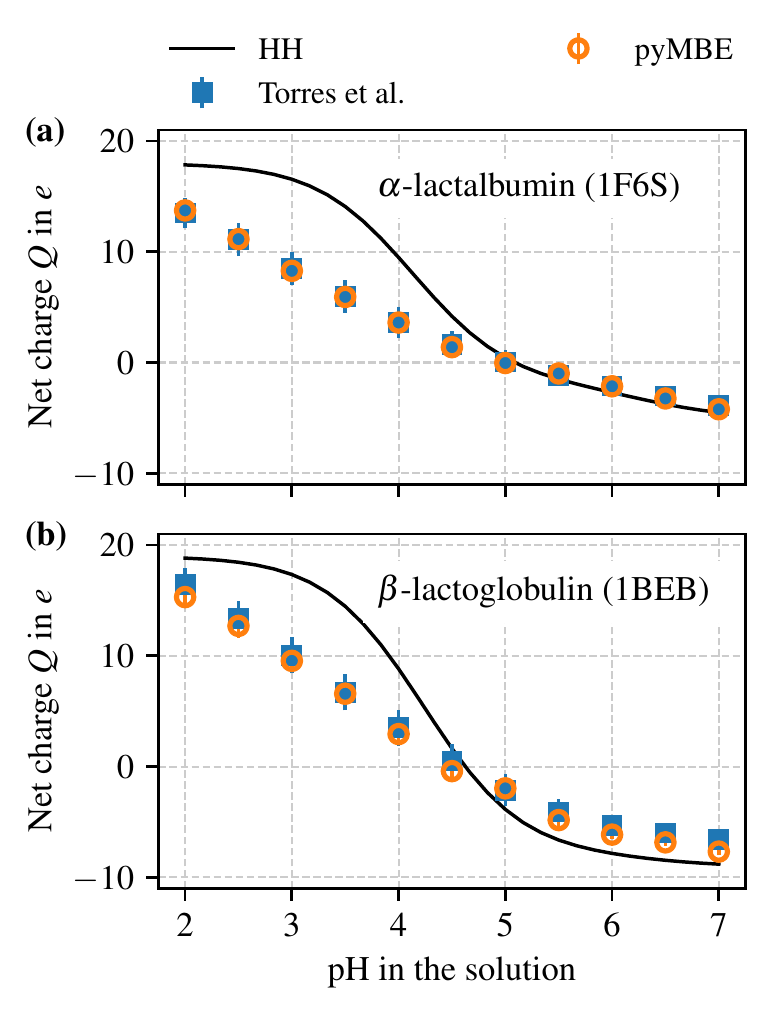}    
    \caption{Net charge $Q$ of $\alpha-$lactalbumin (panel a) and  $\beta-$lactoglobulin (panel b) as a function of the pH.
    The \pymbe module has been used to set up a coarse-grained model of each protein from their PDB crystallographic data: $\alpha-$lactalbumin (PDB code: 1F6S\cite{chrysina2000}) and  $\beta-$lactoglobulin (PDB code: 1BEB \cite{brownlow1997}).
    The net charge measured using \pymbe (orange circles) matches the reference data reported by Torres \etal\cite{torres19a,torres22a} (blue squares) within the estimated error.
    For reference, the analytical solution of the Henderson--Hasselbalch (HH) equation Eqs. \ref{eq:alpha_HH}-\ref{eq:net_charge} for each protein is plotted as a black line.    }
    \label{fig:proteins-netcharge}
\end{figure}

Simulations of rigid globular proteins represent another example of the advantages of using \pymbe to create complex molecules, composed of many sub-units of different types and acidity.
To demonstrate this, we used two globular proteins most abundant in milk whey: $\alpha-$lactalbumin and $\beta-$lactoglobulin.
The acid-base properties of these proteins were simulated by Torres \etal\cite{torres19a, torres22a}
They were represented as rigid objects with two beads per amino acid residue, using the crystallographic structures available in the Protein Data Bank (PDB). 
The original simulations were performed using an in-house Monte Carlo software developed in their group.\cite{torres19a, torres22a}
We used the \pymbe functions for creating the models of globular proteins from their PDB structures, as described in Section \ref{sec:pre-defined}.
In the case of $\alpha-$lactalbumin and additional charge of $+2e$ is included to account for the $\ce{Ca^{2+}}$ ion caged in the protein structure.
In Fig. ~\ref{fig:proteins-netcharge}, we show that constant-pH simulations in \espresso reproduced the net charge $Z$ as a function of pH, reported in the original study by Torres \etal,\cite{torres19a, torres22a} obtained using their own Monte Carlo software.

\subsection{Dialysis of weak polyelectrolyte chains} 
\label{sec:case_study_weak_pe}
\begin{figure}
\centering
\includegraphics[width=\columnwidth]{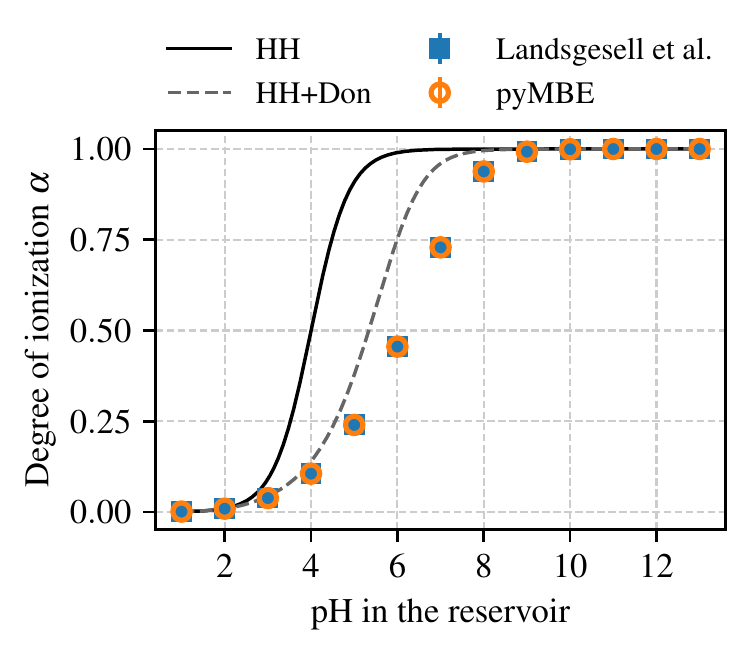}
\caption{\label{fig:dialysis_landsgesell} Degree of ionization $\alpha$ of a weak polyacid solution coupled to a reservoir vs. the pH in the reservoir, obtained for a salt concentration of $c_{\ce{NaCl}}^{\mathrm{res}}=10\,\mathrm{mM}$ in the reservoir and a monomer concentration of $c_{\mathrm{mon}}=435\,\mathrm{mM}$ in the solution. 
The reference data by Landsgesell \etal\cite{landsgesell20b} was also obtained using \espresso. 
"HH" corresponds to the result calculated using the ideal Henderson--Hasselbalch equation (Eq. \ref{eq:alpha_HH}), while "HH+Don" results from a coupled systems of equations involving the Henderson--Hasselbalch equation and the ideal Donnan theory.\\}
\end{figure}

As our final benchmark, we present a simulation of an acid-base equilibrium in a two-phase system that requires the grand-reaction (\grxmc) method in order to correctly model the interplay of charge regulation and ion partitioning.
The model system consists of a solution of weak polyacid chains which is separated by a semi-permeable membrane from an aqueous solution of small ions, termed \textit{reservoir}. 
This system has been used by Landsgesell \etal\cite{landsgesell20b} to validate the \mbox{\grxmc} method.
In brief, the polyelectrolyte is represented by a bead-spring polymer model derived from the Kremer--Grest model.\cite{grest86a}
The simulated system contains 16 chains, each composed of $N=50$ weakly acidic monomers with $\text{p}K_\text{A}=4.0$.
The simulations are performed at different reservoir compositions (salt concentration and pH) and concentrations of the polyelectrolyte within the system.
Full description of the model and its parameters can be found in Ref.\cite{landsgesell20b}

The  \grxmc simulation requires setting up of multiple chemical reactions which are mutually coupled.
Some of these reactions represent the acid-base ionization, others represent the exchange of ion pairs between the system and the reservoir.
To ensure that the simulation works correctly and efficiently in a broad range of pH and salt concentrations, additional reactions should be set up, obtained as linear combinations of the previous ones.
Because the number of reactions is greater than the number of independent parameters, an error in the setup may result in unphysical behavior of the system.
The use of \pymbe in combination with \espresso facilitates the setup of such a system, helping the user avoid common errors.

In Fig. \ref{fig:dialysis_landsgesell}, we plot the degree of ionization $\alpha$ of the polyacid chains as a function of the pH in the reservoir, at a monomer concentration inside the system of $c_{\mathrm{mon}}=435\,\mathrm{mM}$ and salt concentration in the reservoir of $c_{\ce{NaCl}}^{\mathrm{res}}=10\,\mathrm{mM}$.
Unlike the previous benchmark results, Fig. \ref{fig:dialysis_landsgesell} contains two sets of reference curves.
The solid line represents the ideal Henderson--Hasselbalch equation, using the pH in the reservoir as an input, and disregarding the Donnan partitioning of \ce{H+} ions.
It was obtained using the \code{calculate_HH} function from \pymbe.
The dashed line represents the ideal Henderson--Hasselbalch prediction, using the pH in the system as an input, \ie accounting for the Donnan partitioning of ions (HH+Don).
It was obtained using the \code{calculate_HH_Donnan} function of \pymbe.
The key feature of the \grxmc method is that it accounts for both, Donnan partitioning and direct electrostatic interactions between the particles.
Fig. \ref{fig:dialysis_landsgesell} shows that our simulation results, obtained with \espresso and using \pymbe to automatically set up all the required reactions, quantitatively agree with the reference results.

\section{Conclusions and Outlooks}
We presented the Python-based Molecule Builder for \espresso (\pymbe).
\pymbe is an open-source software that facilitates the setup of coarse-grained (CG) models of complex molecules in the \espresso simulation software.
It provides a user-friendly way of building models of flexible peptides and disordered proteins, as well as rigid globular proteins.
Additionally, it facilitates an automated setup of acid-base reactions for constant-pH simulations, as well as exchange of small ions between two phases in simulations using the grand-reaction method.
The use of \pymbe reduces the risk of user errors when setting up such complex simulations, thereby lowering the barrier for new users to start using these methods, implemented in \espresso simulation software.
The development of a common tool to build molecules and set up chemical reactions in such simulations represents also an important step towards reproducibility of results.

In a set of benchmark results, we showed that simulation setups using \pymbe reproduced reference data from constant-pH and grand-reaction simulations produced by different research groups and software.
In addition to benchmarking, these reference results will also serve as a set of test cases for future development of the \pymbe module.

The \pymbe module currently supports only a limited set of interaction potentials (force fields).
These force fields have been selected to enable reproducing the  reference results mentioned above.
Therefore, they may not necessarily represent the most common set of force fields used for modeling of peptides and proteins.
However, thanks to \pymbe, replacing a force field or using a different set of interaction parameters becomes a straightforward task.

The \pymbe module is still under active development as a collaborative project between researchers in the field of physics and chemistry of weak polyelectrolytes and biomacromolecules. 
We plan to continue developing the \pymbe module by extending it to build CG models of other molecules, such as polymer networks, dendrimers and other branched structures and nanoparticles.
We also plan to extend the \pymbe module to include other established force fields for CG models of biomolecules, such as MARTINI\cite{souza2021a}.

\begin{acknowledgments}
We thank the early users of the \pymbe module who have contributed to its development by providing valuable feedback on their experience when using the library: 
Marius Aarsten,
Corinna Dannert,
Rita S. Dias,
Sergio Madurga,
Alberto Martinez-Serra,
Francesc Mas,
Magdaléna Nejedlá,
Cristina Landa Barrio and
Raju Lunkad.

D.B. acknowledges the German Research Foundation (DFG) for funding within the Research Unit FOR2811 ``Adaptive Polymer Gels with Model Network Structure'' under grant 423791428 along with grant 397384169 (TP7).
P.B.T. acknowledges Ph.D. fellowship from CONICET.
S.P. and P.K. acknowledge financial support of the Czech Science foundation, grant 21-31978J and grant agency of the Charles University, GAUK project No. 150224.
C.F.N. acknowledges the financial support from AGENCIA I+D+i (FONCyT), PICT-2021-GRFTI-00090 and from Universidad Tecnológica Nacional (PIDs PATCASR0008459 and PATCASR0008463).
\mbox{J.-N.G.} acknowledges the DFG for funding under grant 528726435 (PI: Holm) and funding by the European Union; this work has received funding from the European High Performance Computing Joint Undertaking (JU) and countries participating in the project under grant agreement No 101093169.
P.M.B. acknowledges the financial support from  the Spanish Ministry of Universities (Margarita Salas Grant MS98), from the Generalitat de Catalunya (Grant 2021SGR00350) and from the European Union’s Horizon Europe research and innovation program under the Marie Skłodowska-Curie grant agreement No 101062456 (ModEMUS).

\end{acknowledgments}

\section*{Data Availability Statement}
The original data showcased in this article in Figs. \ref{fig:peptide_tests}-\ref{fig:dialysis_landsgesell} is available in the \pymbe repository in the folder \mbox{`\texttt{/testsuite/data/src'}} with the permission of its original authors. 
These data sets were originally published in Refs.\cite{landsgesell20b,blanco21a,lunkad21a,torres19a, torres22a}
We also provide the set of scripts that we used to reproduce the data of these articles with \pymbe in the repository of the software in the folder \mbox{`\texttt{/samples/Beyer2024'}}. These scripts are part of pyMBE release 0.8.0\cite{pymbe-zenodo-0.8.0} available on Zenodo.

\section*{Conflict of Interest}
\noindent The authors have no conflicts to disclose.

\section*{Author Contributions}

D.B.: Software, Validation, Visualization, Writing – original draft, Writing – review \& editing.
P.B.T.: Software, Validation, Visualization, Writing – original draft. 
S.P.P.: Software, Visualization, Writing – original draft.
C.F.N.: Funding acquisition, Supervision, Writing – review \& editing.
\mbox{J.-N.G.}: Software, Writing – review \& editing.
P.K.: Conceptualization, Funding acquisition, Supervision, Writing – review \& editing.
P.M.B: Conceptualization, Funding acquisition, Project administration, Software, Supervision, Validation, Visualization, Writing – original draft, Writing – review \& editing.

\appendix

\section{\label{sec:dependencies}Dependencies of \pymbe}
Currently, the \pymbe module depends on the following libraries: \espresso \cite{weik19a,weeber24a}, 
NumPy \cite{harris20a}, 
Pandas \cite{mckinney2010a,pandas2023}, 
Pint \cite{pint}, 
Pint-Pandas \cite{pint-pandas23a} 
and, only for developers, 
pdoc \cite{pdoc}.
Continuous testing and documentation generation are automated using EESSI\cite{droge23a} via GitHub Action eessi/github-action-eessi \cite{eessi-github-action2024}.
We note that these dependencies apply to the current version of \pymbe at the time of writing and they
might change in the future as \pymbe is a live project.
We therefore recommend the reader to visit the repository of \pymbe for up-to-date information about the dependencies of the \pymbe module.

\section{\label{sec:appendix_pymbe_df} Additional examples of the \pymbe Dataframe}
To extend the description of the \pymbe dataframe that we provided in Section \ref{sec:pymbe_df}, let us consider one of the example cases presented in Section \ref{sec:custom}.
For simplicity, let us consider a user that uses the \pymbe module to create two molecules of the type \mbox{\code{name = "alternating_polymer"}} with a small number of monomeric units, for example two.
When creating objects in \espresso, \pymbe adds one row per object into the \pymbe dataframe to bookkeep the information of every individual object created.
For example, if the user filters the \pymbe dataframe by \mbox{\code{pmb_type = "residue"}}, \pymbe returns the dataframe depicted in Table \ref{tab:df_example_res}. 
The first two rows correspond to residues belonging to a molecule with \code{molecule_id = 0} while the last two  correspond to a second molecule with a \code{molecule_id = 1}.
The user can also filter by \code{pmb_type = "bond"} to check how \pymbe has bonded the particles to create the two molecules in the system as shown in Table \ref{tab:df_example_bond}. 
In this case, the subset of the \pymbe dataframe displays six rows, corresponding to the number of bonds needed to build the two example molecules.
Numerical identifiers of the two bonded particles within the \espresso system are stored in \code{particle_id1} and \code{particle_id2}.
The \code{bond_object} stores the instance of the bond object from the \espresso library used to set up the bonding potential between the particle, in which the parameters of the bonding potential can be consulted.
If the user filters the \pymbe dataframe by the \code{pmb_type = "molecule"}, \pymbe returns the dataframe depicted in Table \ref{tab:df_example_mol}.
Each row of the dataframe corresponds to a different molecule in the \espresso system, where the user can observe that the molecules with \code{molecule_id} of 0 and 1 correspond of molecules of the type given by  \code{name = "alternating_polymer"}.

\begin{table}[!htb]
\caption{\label{tab:df_example_res} 
Example of a subset of the \pymbe dataframe, filtered by \code{pmb_type = "residue"}. 
}
\resizebox{\columnwidth}{!}
{%
\begin{tabular}{ccccccc}
\toprule
{\textbf{name}} & {\textbf{pmb\_type}} & {\textbf{residue\_id}} & {\textbf{molecule\_id}} & {\textbf{central\_bead}} & {\textbf{side\_chains}} & {$\cdots$}  \\ \midrule
IA & residue & 0 & 0 & I & [I,A]    & $\cdots$ \\\addlinespace[0.4em]
IB & residue & 1 & 0 & I & [I,B] & $\cdots$ \\\addlinespace[0.4em]
IA & residue & 2 & 1 & I & [I,A]    & $\cdots$  \\\addlinespace[0.4em]
IB & residue & 3 & 1 & I & [I,B]    & $\cdots$  \\\addlinespace[0.4em]
$\vdots$ & $\vdots$ & $\vdots$ & $\vdots$ & $\vdots$ & $\vdots$  & $\ddots$ \\ \bottomrule
\end{tabular}%
}
\end{table}

\begin{table}[!htb]
\caption{\label{tab:df_example_bond} Example of a subset of the \pymbe dataframe, filtered by \code{pmb_type = "bond"}. 
}
\resizebox{\columnwidth}{!}
{%
\begin{tabular}{cccccc}
\toprule
{\textbf{name}} & {\textbf{pmb\_type}} & {\textbf{particle\_id}} & {\textbf{particle\_id2}} & {\textbf{bond\_object}}  & {$\cdots$}  \\ \midrule
I-A & bond & 0 & 1  & HarmonicBond() & $\cdots$ \\\addlinespace[0.4em]
I-I & bond & 0 & 2  & HarmonicBond() & $\cdots$ \\\addlinespace[0.4em]
I-B & bond & 2 & 3  & HarmonicBond() & $\cdots$  \\\addlinespace[0.4em]
I-A & bond & 2 & 4  & HarmonicBond() & $\cdots$  \\\addlinespace[0.4em]
I-I & bond & 4 & 5  & HarmonicBond() & $\cdots$ \\\addlinespace[0.4em]
I-B & bond & 4 & 6  & HarmonicBond() & $\cdots$ \\\addlinespace[0.4em]
$\vdots$ & $\vdots$ & $\vdots$ & $\vdots$ & $\vdots$   & $\ddots$ \\ \bottomrule
\end{tabular}%
}
\end{table}

\begin{table}[!htb]
\caption{\label{tab:df_example_mol} Example of a subset of the \pymbe dataframe, filtered by \code{pmb_type = "molecule"}. 
}
\resizebox{\columnwidth}{!}
{%
\begin{tabular}{ccccc}
\toprule
{\textbf{name}} & {\textbf{pmb\_type}} & {\textbf{molecule\_id}} & {\textbf{residue\_list}} & {$\cdots$}  \\ \midrule
alternating\_polymer & molecule & 0   & [IA,IB]  & $\cdots$ \\\addlinespace[0.4em]
alternating\_polymer & molecule & 1   & [IA,IB]  & $\cdots$ \\
$\vdots$ & $\vdots$ & $\vdots$ & $\vdots$ & $\ddots$ \\ \bottomrule
\end{tabular}%
}
\end{table}

\section{\label{sec:reservoir_composition_grxmc} Reservoir composition in the G-RxMC method}

As explained in Section \ref{sec:GRxMC}, composition of the reservoir in the grand-reaction method is specified \textit{via} the chemical potentials of ions in the reservoir.
These ions are exchanged between the system and reservoir as electroneutral ion pairs.
Within the reaction ensemble framework, these exchanges are represented as virtual chemical reactions, \refeq{eq:insertion_water}--\refeq{eq:insertion_hcl}.
The acceptance probabilities of these virtual chemical reactions are the same as the exchange of ion pairs in the grand canonical ensemble.\cite{landsgesell20b}
The equilibrium constant for each ion pair is related to the sum of their chemical potentials in the reservoir:
\begin{align}
    \kT \ln K_{ij} = \mures_i + \mures_j - \muref_i - \muref_j
    \label{eq:lnKij}
\end{align}
Furthermore, the value of $K_{ij}$ is related to the concentrations of these ions in the reservoir
\begin{align}
    K_{ij} = \frac{\gamma_{\pm}^2 \cres_i \cres_j}{(\cref)^2}
    \label{eq:Kij}
\end{align}
where $\gamma_{\pm} = (\exp(\beta \muex_i) + \exp(\beta \muex_j) ) / 2$ is the mean activity coefficient at the given composition of the reservoir and $\muex_i$ is the excess chemical potential of ion $i$.
In an ideal system, $\gamma_{\pm} = 1$, and the constants $K_{ij}$ are simply given by the products of concentrations of individual ions.
The self-ionization of water is given by its equilibrium constant at room temperature
\begin{align}
\Kw = \gamma_{\pm}^2 c_{\ce{H+}} c_{\ce{OH-}} / (\cref)^2 = 10^{-14}
    \label{eq:Kw}
\end{align}
Out of the three remaining constants, only two can be chosen independently, because the last one is determined by the requirement of electroneutrality in the reservoir.
In a non-ideal system containing only monovalent ions, $\gamma_{\pm}$ depends on the ionic strength of the solution, 
\begin{align}
    I = \frac{1}{2} \sum_i z_i^2 c_i
    \label{eq:I}
\end{align}
where the summation runs over all small ions present in the reservoir.
Therefore, constants $K_{ij}$ depend not only on the concentrations of individual ions but also on the ionic strength.

In contrast with simulations in the grand-reaction ensemble, the reservoir composition in experiments would be typically specified by the concentrations of individual ionic species and by pH in the reservoir, defined as
\begin{align}
    \mathrm{pH}^\mathrm{res} 
    = -\log_{10}\left(\frac{\gamma_{\pm} \cres_{\ce{H+}}}{\cref}\right)
    = -\frac{\mures_{\ce{H+}} - \muref_{\ce{H+}} }{\ln 10}
    \label{eq:pH-def}
\end{align}
For simplicity, we assume that the reservoir contains only monovalent salt ions, which we call \ce{Na+} and \ce{Cl-}, and the \ce{H+} and \ce{OH-} ions, which are inevitably present in aqueous solutions.
This could be prepared by dissolving a given amount of NaCl in pure water and then adjusting the pH by adding extra NaOH or HCl.
Because \pymbe currently implements all ions as point charges with WCA potential for steric repulsion, the discussion below applies to any reservoir which contains only monovalent ions represented in the same way.

When the ionic strength of the solution is determined predominantly by the NaCl concentration, one can use for example the extended Debye--Hückel theory to estimate the activity coefficients from the ionic strength,\cite{atkins23a} 
\begin{align}
    -\log_{10} \gamma_{\pm} = \frac{ A\sqrt{I} }{1 + Ba\sqrt{I} }
    \label{eq:gamma-DH}
\end{align}
where $a$ is the effective diameter of the ion in nanometers and $A$ and $B$ are constants, whose values for aqueous solutions at room temperature are $A = \qty{0.5085}{L^{1/2} mol^{-1/2}}$ and $B = \qty{0.3281}{ L^{1/2}  mol^{-1/2} nm^{-1}}$.
The calculation becomes more complex at extreme pH values, when the amount of added NaOH or HCl exceeds the amount of NaCl.
In the general case, the activity coefficients (excess chemical potentials) are uniquely determined by the ionic strength, which depends on the concentrations of all ions
\begin{align}
    \gamma_{\pm}(\Ires) = f(\Ires) = f\left(\Ires \left(c_{\ce{H+}}^\mathrm{res},c_{\ce{OH-}}^\mathrm{res},c_{\ce{NaCl}}^\mathrm{res}\right)\right),
\label{eq:gamma-general}
\end{align}
At low ionic strength, $f$ can be approximated by \refeq{eq:gamma-DH}, while at high ionic strength, it can be estimated from simulations using the Widom insertion method.\cite{widom63a}
As explained in Section \ref{sec:GRxMC}, in pyMBE the functional form of $f$ is not fixed but has to be provided by the user when setting up the G-RxMC method.
In the general case, Eq. \ref{eq:gamma-general}, \refeq{eq:pH-def} and \refeq{eq:Kw} form a coupled system of nonlinear equations that can be solved numerically for the concentrations and the excess chemical potential, given values for $\mathrm{pH}^\mathrm{res}$ and $c_{\ce{NaCl}}^\mathrm{res}$.
In \pymbe, this calculation is performed iteratively until self-consistency.

\section{\label{sec:HH_coupled_Donnan} Henderson--Hasselbalch equation coupled to the Donnan equilibrium}

In reactive two-phase systems, charge regulation and ion partitioning are inherently coupled in a highly non-linear fashion, as shown in Fig.\ref{fig:schematic_ensembles}(b). 
For ideal systems, this leads to a system of equations which can be solved numerically to obtain the Donnan potential and ionization degrees under the given conditions. 
As in the case of single-phase systems, this ideal description is often an important reference point to assess the role of (electrostatic) interactions. 

In an ideal model, the partitioning of small ions is described by the Donnan theory, and the charge regulation of the weak acid and base groups, is described by the Henderson--Hasselbalch equation. 
The partitioning of exchangeable ionic species is determined by the equality of electrochemical potentials,
\begin{align}
 \mu_n^\mathrm{sys} + q_n\psi^\mathrm{Don} = \mu_n^\mathrm{res}
 \label{eq:electrochemical_equilibrium}
\end{align}
where $n\in\{\ce{H+}$, $\ce{OH-}$, $\ce{Na+}$, $\ce{Cl-}\}$.
Here, $q_n$ is the charge of ion $n$ and $\psi^\mathrm{Don}$ is the Donnan potential, which plays the role of a Lagrange multiplier that ensures the electroneutrality of both phases.\cite{landsgesell20b} 
Consequently, the resulting equilibrium is also often termed the "Donnan equilibrium".
Assuming an ideal system, \ie neglecting all interactions, the Donnan potential can be related to the partition coefficient of cations:
\begin{align}
\xi_+ 
    \equiv \frac{c_+^\text{sys}}{c_+^\text{res}}
    = \exp\left(-\beta e\psi^\mathrm{Don}\right) 
\end{align}
and a closed expression for the Donnan potential can be obtained:\cite{landsgesell20b}
\begin{align}
\exp\left(-\beta e\psi^\mathrm{Don}\right) 
    = \frac{\sum_i z_i\alpha_i c_i}{2I^\mathrm{res}} + \sqrt{\left(\frac{\sum_i z_i\alpha_i c_i}{2I^\mathrm{res}}\right)^2+1}.
 \label{eq:donnan_partitioning}
\end{align}
In this equation,
\begin{align}
I^\mathrm{res} \equiv \frac{c_{\ce{H+}}^{\mathrm{res}}+c_{\ce{OH-}}^{\mathrm{res}}+c_{\ce{Na+}}^{\mathrm{res}}+c_{\ce{Cl-}}^{\mathrm{res}}}{2}
\end{align}
is the ionic strength of the reservoir, taking into account the small ions $\ce{H+}$, $\ce{OH-}$, $\ce{Na+}$, $\ce{Cl-}$. 
The sum $\sum_i z_i\alpha_i c_i$ runs over all  ionizable groups in the system, which cannot be exchanged with the reservoir.
More specifically, $z_i$ is the charge number of group $i$ when it is ionized, $\alpha_i$ is the corresponding degree of ionization and $c_i$ the total concentration of groups of type $i$ in the system, irrespective of the ionization state. 
For charge-regulating systems, the degrees of ionization $\alpha_i$ can be calculated from the pH in the system using the Henderson--Hasselbalch equation \refeq{eq:alpha_HH}. 
 However, the pH in the system differs from pH in the reservoir by the Donnan contribution:
 \begin{align}
	\text{pH}^\text{sys} - \text{pH}^\text{res} = \frac{\beta e\psi^\text{Don}}{\ln 10}.
\label{eq:Donnan_effect_pH_appendix}
\end{align}
Under the ideal gas approximation, the degree of ionization $\alpha_i$ is thus given by
\begin{equation}
\alpha_i \ideq \frac{1}{1 + 10^{z_i\left(\pKai - \pH^\text{sys}\right)}} 
\label{eq:alpha_sys}
\end{equation}
where $\pH^\text{sys}$ depends on $\text{pH}^\text{res}$ and $\psi^\text{Don}$ through Eq. \ref{eq:Donnan_effect_pH_appendix}. Equations \ref{eq:donnan_partitioning} and \ref{eq:alpha_sys} form a system of non-linear equations relating the degrees of ionization and the Donnan potential; charge regulation and ion partitioning are thus inherently coupled.
Inserting the equations into each other, it is possible to reduce the set of equations to a non-linear equation for a single variable. 
In practice, it is convenient to formulate this equation in terms of the partition coefficient $\xi_+$:
\begin{align}
\xi_+ 
    = \frac{\sum_i z_i\alpha_i c_i}{2I^\mathrm{res}} + \sqrt{\left(\frac{\sum_i z_i\alpha_i c_i}{2I^\mathrm{res}}\right)^2+1},
\end{align}
where
\begin{align}
\alpha_i = \frac{1}{1 + 10^{z_i\left(\pKai - \pH^\text{res} + \mathrm{log}_{10} \xi_+\right)}}.
\end{align}
While not solvable analytically, this equation can be solved numerically using standard root finding techniques.
In the \pymbe module, this calculation can be performed using the function \code{calculate_HH_Donnan}, as demonstrated in Section \ref{sec:case_study_weak_pe}.

\bibliography{Bibliography.bib}
\end{document}